\documentclass[10pt, sigconf, twocolumn]{acmart}
\usepackage{graphicx}
\usepackage{balance}  

\usepackage{afterpage}
\usepackage{pifont}
\usepackage{float}
\usepackage{amsmath}
\usepackage{listings}
\usepackage{empheq}
\usepackage[framemethod=tikz,xcolor=true]{mdframed}

\usepackage{amsthm}
\usepackage{amsfonts}
\usepackage{cleveref}
\usepackage{colortbl}
\usepackage{paralist}
\usepackage{enumitem}
\usepackage{epsfig}
\usepackage{forest}
\usepackage{graphicx}
\usepackage{multicol}
\usepackage{multirow}
\usepackage{pgf}
\usepackage{pdflscape}
\usetikzlibrary{positioning}
\usepackage{rotating}
\usepackage[normalem]{ulem}
\usepackage{relsize}
\usepackage{textcomp}
\usepackage{tikz}
\usepackage{wrapfig}
 \usepackage{xspace}
\usepackage{balance}
\usepackage{thmtools,thm-restate}
\usepackage[normalem]{ulem}
\usepackage{url}
\usepackage{amsmath}
\usepackage{amssymb}
\usepackage{xcolor}
\usepackage{mathtools}
\usepackage{subcaption}
\usepackage{hyperref}
\usepackage{xpatch}
\usepackage[vlined, ruled, linesnumbered]{algorithm2e}

\DeclarePairedDelimiter\ceil{\lceil}{\rceil}

\definecolor{awesome}{rgb}{1.0, 0.13, 0.32}
\definecolor{bittersweet}{rgb}{1.0, 0.44, 0.37}

\renewcommand{\paragraph}[1]{\vspace{2pt} \noindent {\bf #1}}

\newcommand{\eat}[1]{}
\newcommand{\later}[1]{}

\newcommand{\system}{\textsc{Cleo}\xspace}

\newcommand{\tar}[1]{\textcolor{violet}{[Tarique: #1]}}

\newcommand{\new}[1]{\textcolor{teal}{[NEW: #1]}}
\newcommand{\todo}[1]{\textcolor{red}{[TODO: #1]}}
\newcommand{\done}[1]{}
\newcommand{\hide}[1]{}
\newcommand{\fb}[1]{\textcolor{red}{[#1]}}

\newcommand{\techreport}[1]{#1}
\newcommand{\papertext}[1]{}

\newcommand{\microsoft}[1]{Microsoft}
\newcommand{\scope}[1]{SCOPE}

\newcommand{\rev}[1]{#1}

\newcommand{\shep}[1]{{{#1}}}

    {\end{list}}

\newcounter{enum}
\newenvironment{packed_enum}{
\begin{list}{\textbf{(\arabic{enum})}}{
  \setlength{\itemsep}{0pt}
  \setlength{\parskip}{0pt}
  \setlength{\labelwidth}{-4 pt}
  \setlength{\leftmargin}{0 pt}
  \setlength{\itemindent}{0pt}
  \usecounter{enum}}
}{\end{list}}

\newcommand{\stitle}[1]{\vspace{0.25em}\noindent\textbf{#1}}

\SetKwProg{Fn}{Function}{}{}

\eat{
\vldbTitle{Learned Cost Models for Optimizing Big Data Queries}
\vldbAuthors{Tarique Siddiqui, Alekh Jindal, Shi Qiao, Hiren Patel, Wangchao Le}
\vldbDOI{https://doi.org/10.14778/xxxxxxx.xxxxxxx}
\vldbVolume{}
\vldbNumber{xxx}
\vldbYear{}
}

\copyrightyear{2020}
\acmYear{2020}
\setcopyright{acmcopyright}
\acmConference[SIGMOD'20]{Proceedings of the 2020 ACM SIGMOD International Conference on Management of Data}{June 14--19, 2020}{Portland, OR, USA}
\acmBooktitle{Proceedings of the 2020 ACM SIGMOD International Conference on Management of Data (SIGMOD'20), June 14--19, 2020, Portland, OR, USA}
\acmPrice{15.00}
\acmDOI{10.1145/3318464.3380584}
\acmISBN{978-1-4503-6735-6/20/06}

\begin{document}


\title{{\huge \system}: Learned Cost Models for Query Optimization in Shared Clouds}
\title{Learned Cost Models for Cloud Data Services}
\title{Learned Cost Models for Query Optimization in the Cloud}
\title{{\huge \system}: Accurate Cost Models for Query Optimization}
\title{Learned Cost Models for Optimizing Big Data Queries}

\title{Cost Models for Serverless Query Processing: Learning, Retrofitting, and Our Findings}
\title{\rev{CLEO: Learned Cost Models for Query Optimization in Big Data Systems}}
\title{\rev{CLEO: Learned Cost Models for Cloud-based Big Data Systems}}
\title{\rev{Cost Models for Big Data Query Processing: Learning, Retrofitting, and Our Findings}}



%
%
%
%


\author{
Tarique Siddiqui$^{1,2}$, Alekh Jindal$^{1}$, Shi Qiao$^{1}$, Hiren Patel$^{1}$, Wangchao Le$^{1}$ \\
$^1$Microsoft\hspace{0.5cm}$^2$University of Illinois, Urbana-Champaign 
}

\begin{abstract}
\rev{Query processing over big data is ubiquitous in modern clouds, where the system
takes care of picking both the physical query execution plans \emph{and} the resources needed to run those plans}, using a cost-based query optimizer. A good cost model, therefore, is akin to better resource efficiency and lower operational costs.
Unfortunately, the production workloads at \microsoft{} show that costs are very complex to model for big data systems.
In this work, we investigate two key questions: 
(i)~can we {\it learn} accurate cost models for big data systems, and 
(ii)~can we {\it integrate} the learned models within the query optimizer.
To answer these, we make three core contributions.
First, we exploit workload patterns to learn a large number of individual cost models and combine them to achieve high accuracy and coverage over a long period.
Second, we propose extensions to Cascades framework to pick optimal resources, i.e, number of containers, during query planning.
And third, we integrate the learned cost models within the Cascade-style query optimizer of \scope{} at \microsoft{}.
We evaluate the resulting system, \system, in a production environment using both production and TPC-H workloads.
Our results show that the learned cost models are $2$ to $3$ orders of magnitude more accurate, and $20\times$ more correlated with the actual runtimes, with a large majority  ($70\%$)  of the plan changes leading to substantial improvements in latency as well as resource usage. 
\end{abstract}

\maketitle
\setcounter{page}{1}
\setcounter{section}{0}
\renewcommand\thefigure{\arabic{figure}}
\setcounter{figure}{0}

\vspace{-6pt}
\section{Introduction}
\label{sec:intro}


\hide{\rev{Modern cloud} computing is changing the way users interact with public clouds,
by having them focus on their computing tasks \emph{without} worrying about provisioning resources for those tasks.
For big data analytics, this means that users specify their declarative queries and a cloud service takes care of 
provisioning the right set of resources (e.g., number of containers) for those queries, e.g., Athena~\cite{aws-athena}, ADLA~\cite{adla}, BigSQL~\cite{ibm-bigsql}, and BigQuery~\cite{google-bigquery}.
Behind the scenes, for each declarative user query, a cost-based optimizer picks the physical execution plan and the resources needed to run that plan.
An accurate cost model is therefore crucial to optimizing the resources used, and hence the operational costs incurred, in \rev{these big data} environments.
}

\begin{figure}
	\centerline {
		\hbox{\resizebox{\columnwidth}{!}{\includegraphics{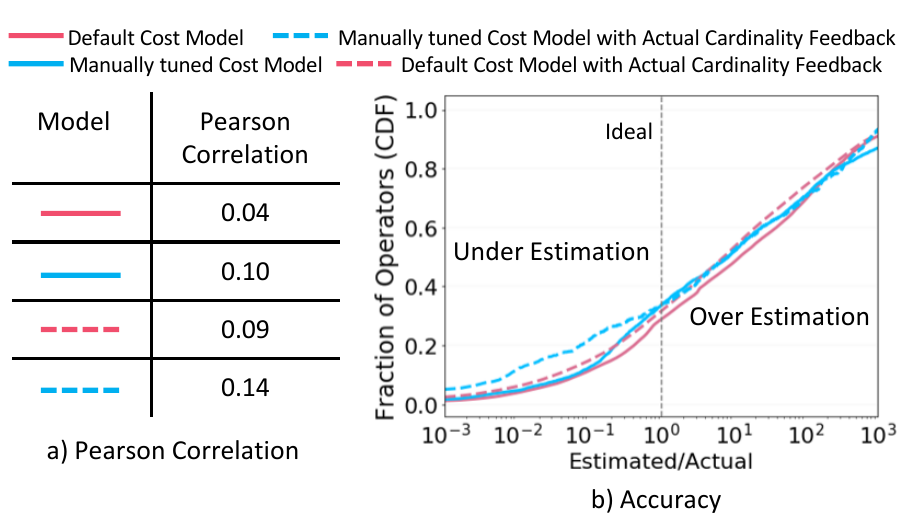}}}}
	\vspace{-14pt}
	\caption{\small Impact of manual tuning and cardinality feedback on cost models in \scope{}}
	\vspace{-15pt}
	\label{fig:costModelwithCardFeedback}
\end{figure}

There is a renewed interest in cost-based query optimization in big data systems, particularly in modern cloud data services (e.g., Athena~\cite{aws-athena}, ADLA~\cite{adla}, BigSQL~\cite{ibm-bigsql}, and BigQuery~\cite{google-bigquery}) that are responsible for picking both the query execution plans and the resources (e.g., number of containers) needed to run those plans.
Accurate {\it cost models} are therefore crucial for generating efficient combination of plan and resources.
Yet, the traditional wisdom from relational databases is that cost models are less important and fixing cardinalities automatically fixes the cost estimation~\cite{leis2015good, lohman2014query}.
The question is whether this also holds for the new breed of big data systems.
To dig deeper, we analyzed one day's worth of query logs from the \rev{big data} infrastructure (\scope{}~\cite{scopeVLDB2008,scopeVLDBJ12}) at \microsoft{}.
We feed back the actual runtime cardinalities, i.e., the ideal estimates that 
 any cardinality estimator, including learned models~\cite{stillger2001leo,cardLearner,lightweightCardModels,corrJoinsCardModels} can achieve.
Figure~\ref{fig:costModelwithCardFeedback} compares the ratio of cost estimates with the actual runtimes for two cost models in \scope{}: 1) a default cost model, and 2) a manually-tuned cost model that is partially available for limited workloads.
The vertical dashed-line at $10^0$ corresponds to an ideal situation where all cost estimates are equal to the actual runtimes. Thus, the closer a curve  is to the dashed line, the more accurate it is.

The dotted lines in Figure~\ref{fig:costModelwithCardFeedback}(b) show that
fixing cardinalities 
reduces the over-estimation, but there is still a wide gap between the estimated and the actual costs, with the Pearson correlation being as low as $0.09$.
This is due to the complexity of big data systems coupled with the variance in cloud environments~\cite{cloudVariance}, which makes cost modeling incredibly difficult.
Furthermore, any improvements in cost modeling need to be consistent across workloads and over time
since performance spikes are detrimental to the reliability expectations of enterprise customers. 
Thus, accurate cost modeling is still a challenge in \scope{} like big data systems. 

\eat{
Cost-based optimizers are crucial to picking efficient execution plans in big data systems~\cite{apache-hive,apache-spark,apache-flink,apache-calcite}. 
The core of cost-based optimizers uses cost models to model operator performances and pick the cheapest plan.
However, with big data systems getting commoditized as data services in the cloud~\cite{aws-athena,scopeVLDB2008,ibm-bigsql,google-bigquery}, cost-based optimization becomes very challenging. This is due to several reasons. First, the complexity of big data systems coupled with the variance in cloud environments~\cite{cloudVariance} makes cost modeling incredibly difficult; inaccurate cost estimates can easily produce plans with significantly poor performance. Second, any improvements in cost estimates need to be consistent across workloads and over time; this makes cost tuning harder since performance spikes are detrimental to the reliability expectations of cloud customers. And finally, reducing total cost of operations is crucial in cloud environments, and so query optimizers need to consider the resources consumed as part of query planning, especially when cloud resources could be provisioned dynamically and on demand.
}


In this paper, we explore the  following two questions:
\vspace{-0.2cm}
\begin{packed_enum}
\item {\bf Can we learn accurate, yet robust cost models for big data systems?}
This is motivated by the presence of massive workloads visible in modern cloud services 
that can be harnessed to accurately model the runtime behavior of queries. 
This helps not only in dealing with the various complexities in the cloud, but also specializing or {\it instance optimizing}~\cite{marcus2019neo} to specific customers or workloads, which is often highly desirable. Additionally, in contrast to years of experience needed to tune traditional optimizers, learned cost models are potentially easy to update at a regular frequency.
\item {\bf Can we effectively integrate  learned cost models within the query optimizer?}
This stems from the observation that while some prior works have considered learning models for predicting query execution times for a given physical plan in traditional databases~\cite{ganapathi2009predicting,bach2002kernel,akdere2012learning,li2012robust}, none of them have integrated learned models within a query optimizer for selecting  physical plans. Moreover, in big data systems, resources (in particular the number of machines) play a significant role in cost estimation~\cite{raqo}, making the integration even more challenging. Thus, we investigate the effects of learned cost models on query plans by \emph{extending} the \scope{} query optimizer in a \emph{minimally invasive} way for \emph{predicting costs} in a \emph{resource-aware manner}. 
\techreport{To the best of our knowledge, this is the first work to integrate learned cost models within an industry-strength query optimizer.}
\end{packed_enum}
\vspace{-0.18cm}

Our key ideas are as follows.
We note that the cloud workloads are quite diverse in nature, i.e., there is no representative workload to tune the query optimizer, and hence there is no single cost model that fits the entire workload, i.e., {\it no-one-size-fits-all}. Therefore, we learn a large collection of smaller-sized cost models, one for each common subexpressions that are typically abundant in production query workloads~\cite{cloudviews,bigsubs}. While this approach results in specialized cost models that are very accurate, the models do not cover the entire workload: expressions that are not common across queries do not have models. The other extreme is to learn a cost model per operator, which covers the entire workload but sacrifices the accuracy with very general models. Thus, \emph{there is an accuracy-coverage trade-off that makes cost modeling challenging}.
To address this, we define the notion of cost model \emph{robustness} with three desired properties: (i) high accuracy, (ii) high coverage, and (iii) high retention, i.e., stable performance for a long-time period before retraining. 
We achieve these properties in two steps:
First, we bridge the accuracy-coverage gap by learning additional mutually enhancing models that improve the coverage as well as the accuracy. Then, we learn a combined model that automatically corrects and combines the predictions from multiple individual models, providing accurate and stable predictions for a sufficiently long window (e.g., more than 10 days). 

We implemented our ideas in a Cloud LEarning Optimizer (\system) and integrated it within \scope{}.
\system uses a feedback loop to periodically train and update the learned cost models within the 
Cascades-style top-down query planning~\cite{cascades95} in \scope{}.
We extend the optimizer to invoke the learned models, instead of the default cost models, to estimate the cost of candidate operators. However, in \rev{big data} systems, the cost depends heavily on the resources used (e.g., number of machines for each operator) by the optimizer~\cite{raqo}. Therefore, we extend the Cascades framework to explore resources, and propose mechanisms to explore and derive optimal  number of machines for each stage in a query plan. Moreover, instead of using handcrafted heuristics or assuming fixed resources, we leverage the learned cost models to find optimal resources as part of query planning, thereby \emph{using learned models for producing both runtime as well as  resource-optimal plans}.

In summary, our key contributions are as follows.

\vspace{-0.12cm}
\begin{packed_enum}
\item We motivate the cost estimation problem from production workloads at \microsoft{}, including prior attempts for manually improving the cost model
(Section~\ref{sec:motivation}).

\item We propose machine learning techniques to learn highly {\it accurate} cost models. Instead of building a generic cost model for the entire workload, we learn a large collection of smaller specialized models that are resource-aware and highly accurate in predicting the runtime costs (Section~\ref{sec:learnedcosts}).

\item We describe the accuracy and coverage trade-off in learned cost models, show the two extremes, and propose additional models to bridge the gap. 
We combine the predictions from individual models into a {\it robust model} that provides the best of both accuracy and coverage over a long period (Section~\ref{sec:robustness}).

\item We describe integrating \system within \scope{}, including periodic training, feedback loop, model invocations during optimization, and novel extensions for finding the optimal resources for a query plan (Section~\ref{sec:opt-intg}). 

\item Finally, we present a detailed evaluation
of \system, using both the production workloads and the TPC-H benchmark.
Our results show that \system improves the correlation between predicted cost and actual runtimes from $0.1$ to $0.75$, the accuracy by $2$ to $3$ orders of magnitude,
and the performance for 70\% of the changed plans (Section~\ref{sec:experiments}).
\papertext{In our technical report~\cite{techreport}, we further describe practical techniques to address performance regressions in our production settings.}
\techreport{In Section~\ref{sec:discussion}, we further describe practical techniques to address performance regressions in our production settings.}
\end{packed_enum}
\vspace{-10pt}

\section{Motivation}
\label{sec:motivation}

In this section, we give an overview of \scope{}, its workload and query optimizer, and motivate the cost modeling problem from  production workloads at \microsoft{}.

\vspace{-0.3cm}
\subsection{Overview of \scope{}}
\label{sec:scopebackgnd}

\scope{}~\cite{scopeVLDB2008,scopeVLDBJ12}  is the big data system used for internal data analytics across the whole of \microsoft{} to analyze and improve its various products.
It runs on a hyper scale infrastructure consisting of hundreds of thousands of machines, running a massive workload of hundreds of thousands of jobs per day that process exabytes of data. \scope{} exposes a job service interface where users submit their analytical queries and the system takes care of automatically provisioning resources and running queries in a distributed environment. 

\eat{
SCOPE~\cite{scopeVLDB2008,scopeVLDBJ12} is the big data system used for internal data analytics across the whole of Microsoft to analyze and improve the various products, including Bing, Office, Skype, Windows, XBox, etc. It runs on a hyper scale infrastructure consisting of hundreds of thousands of machines, running a massive workload of hundreds of thousands of jobs per day that process exabytes of data. SCOPE exposes a job service interface where users submit their analytical queries and the system takes care of automatically provisioning resources and running queries in a distributed environment. 
}

\scope{} query processor partitions data into smaller subsets and processes them in parallel. The number of machines running in parallel (i.e., degree of parallelism) depends on the number of partitions of the input.
When no specific partitioning is required by upstream operators, certain physical operators (e.g., \texttt{Extract} and \texttt{Exchange} (also called \texttt{Shuffle})), decide partition counts based on data statistics and heuristics.  The sequence of intermediate operators that operate over the same set of input partitions are grouped into a \texttt{stage} --- all operators in a stage run on the same set of machines.  Except for selected scenarios, Exchange operator is commonly used to re-partition data between two stages.

\papertext{\vspace{-7pt}}
\subsection{Recurring Workloads}
\label{sec:recurring}
\shep{
SCOPE workloads primarily consist of {\bf recurring jobs}. 
A recurring job in SCOPE is used to provide periodic (e.g., hourly, six-hourly, daily, etc.)  analytical result for a specific application functionality. Typically, a recurring job consists of a script template that accepts different input parameters similar to SQL modules. Each {\it instance} of the recurring job runs on different input data, parameters and have potentially different statements. As a result, each instance is different in terms of input/output sizes, query execution plan, total compute hour, end-to-end latency, etc. Figure~\ref{fig:recurring_job_example} shows $150$ instances of an hourly recurring job that extracts facts from a production clickstream.
Over these $150$ instances, we can see a big change in the total input size and the total execution time, 
from $69,859$ GiB to $118,625$ GiB
and from $40$ mins and $50$ seconds to $2$ hours and $21$ minutes respectively.
Note that a smaller portion of SCOPE workload is ad-hoc as well. 
\papertext{Our analysis from four of the production clusters shows $7\%-20\%$ ad-hoc jobs on a daily basis.}
\techreport{Figure~\ref{fig:non_recurring_jobs} shows our analysis from four of the production clusters. We can see that $7\%-20\%$ jobs are ad-hoc on a daily basis, with the fraction varying over different clusters and different days.}
However, compared to ad-hoc jobs, recurring jobs represent long term business logic with critical value, 
and hence the 
focus of several prior research works~\cite{bruno2013continuous, recurrJobOpt, recurrJobOptScope, redoop, rope, jockey, jyothi2016morpheus, cloudviews, bigsubs, cardLearner}
and also the primary focus for performance improvement in this paper.


%
%
%
%
%
%
%
%
%
%

\begin{figure}
	\vspace{1pt}
	\centerline {
		\hbox{\resizebox{\columnwidth}{!}{\includegraphics{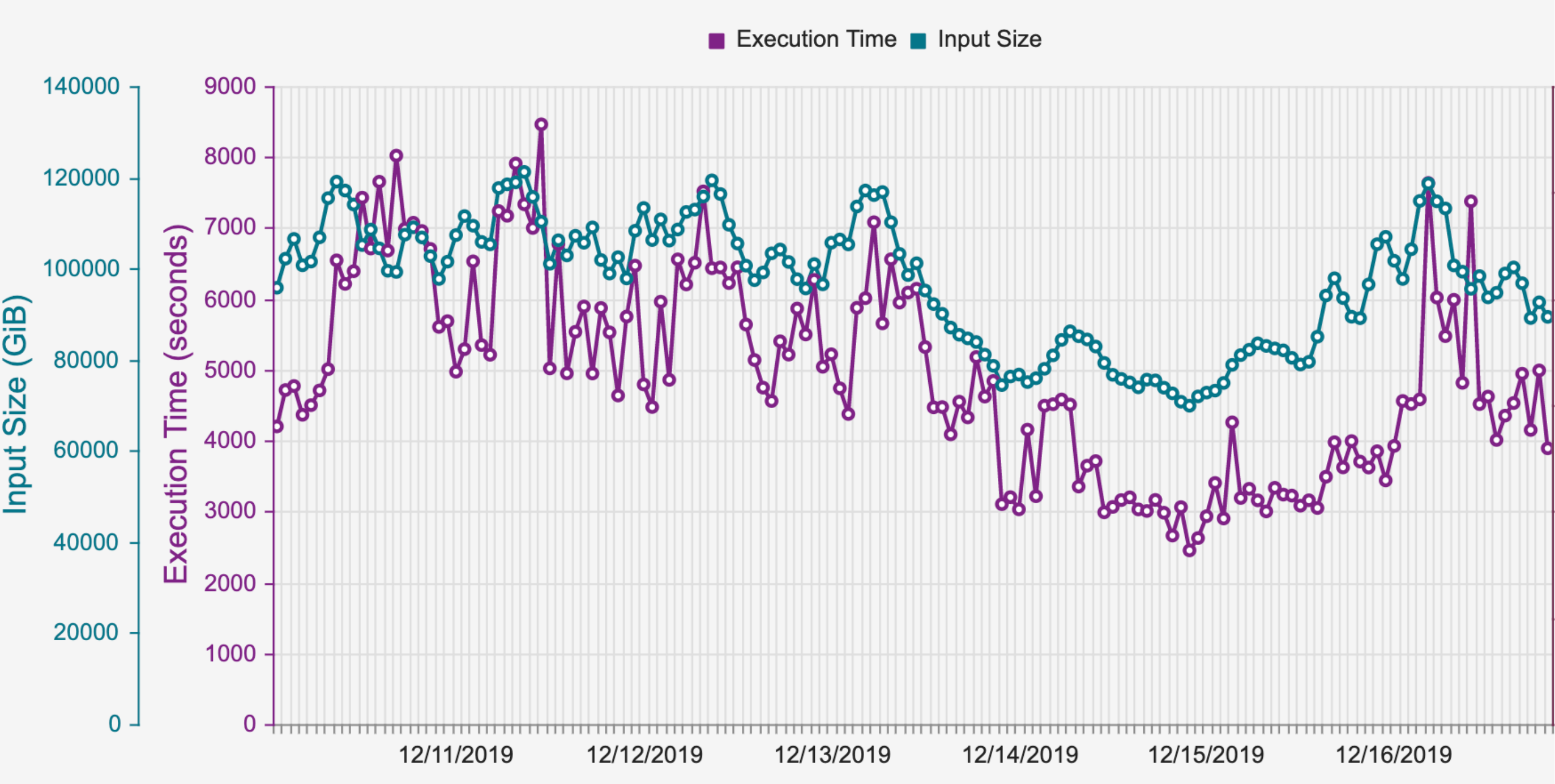}}}}
	\vspace{-5pt}
	\caption{\shep{$150$ instances of an hourly {\it recurring job} that extracts facts from a production clickstream.}}
	\label{fig:recurring_job_example}
	\papertext{\vspace{-15pt}}
\end{figure}

\techreport{
\begin{figure}
	\centerline {
		\hbox{\resizebox{0.75\columnwidth}{!}{\includegraphics{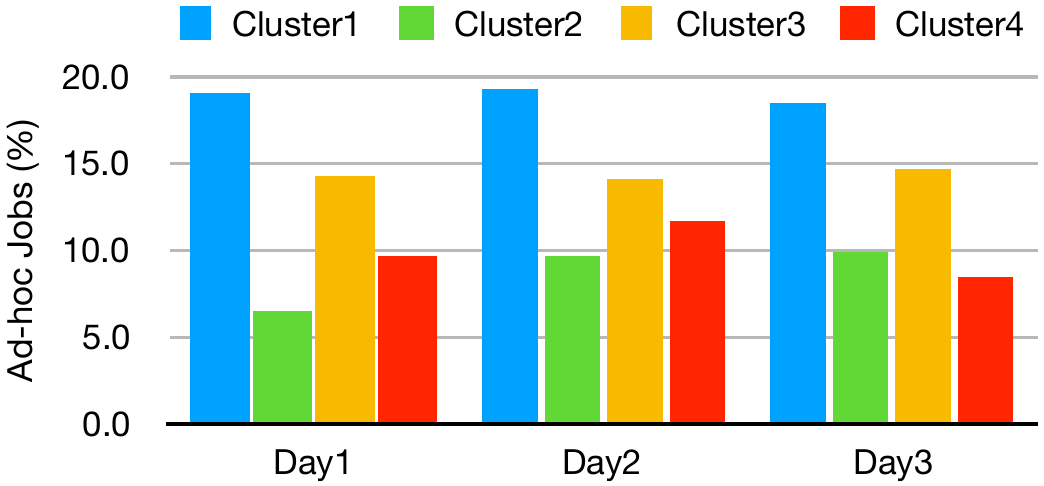}}}}
	\caption{Illustrating ad-hoc jobs in SCOPE. }
	\label{fig:non_recurring_jobs}
	\vspace{-12pt}
\end{figure}
}

}

\papertext{\vspace{-0.25cm}}
\subsection{Overview of SCOPE Optimizer} 
\label{sec:optoverview}
\scope{} uses a cost-based optimizer based on the Cascades Framework~\cite{cascades95} for generating the execution plan for a given query.
Cascades~\cite{cascades95} transforms a logical plan using multiple tasks: (i)~Optimize Groups, (ii)~Optimize Expressions, (iii)~Explore Groups, and (iv)~Explore Expressions, and (v)~Optimize Inputs. While the first four tasks search for candidate plans via transformation rules, our focus in this work is essentially on the Optimize Inputs tasks, where the cost of a physical operator in estimated. The cost of an operator is modeled to capture its {\em runtime latency}, estimated using a combination of data statistics and hand-crafted heuristics developed over many years. Cascades performs optimization in a top-down fashion, where physical operators higher in the plan are identified first. The exclusive (or local) costs of physical operators are computed and combined with costs of children operators to estimate the total cost.
In some cases, operators can have a \emph{required property} (e.g., sorting, grouping) from its parent that it must satisfy, as well as can have a {\em derived property} from its children operators. In this work, we optimize \emph{how partition counts are derived} as it is a key factor in cost estimation for massively parallel data systems. 
Overall, our goal is to improve the cost estimates with minimal changes to the Cascades framework. We next analyze the accuracy of current cost models in \scope{}.

\papertext{\vspace{-0.57cm}}
\subsection{Cost Model Accuracy}
\label{sec:cm_accuracy}

\done{To analyze the accuracy of the \scope{} cost model, we analyzed one day of production workloads from one of the largest internal customers.
Figure~\ref{fig:costModelwithCardFeedback}a (solid red) shows an extremely low Pearson correlation of $0.04$ between the cost estimates produced by the default cost model and the actual runtime latency. We further look at the ratio of the cost estimates and the actual runtime latency in Figure~\ref{fig:costModelwithCardFeedback}~b.
A ratio close to $1$ would mean that the estimated costs are close to the actual ones. However, as we can see from the figure, the estimated and actual cost ratio ranges from $10^{-2}$ to $10^3$, indicating a wide mismatch, with both significant under- and significant over- estimation.}

\rev{The solid red line in Figure~\ref{fig:costModelwithCardFeedback} shows that the cost estimates from the default cost model range 
between an  under-estimate of $100 \times$ to an over-estimate of $1000 \times$, with a Pearson correlation of just $0.04$.}
As mentioned in the introduction, this is because of the difficulty in modeling the highly complex big data systems.
Current cost models rely on hand-crafted heuristics that combine statistics (e.g., cardinality, average row length)  in complex ways to estimate each operator's execution time. These estimates are usually way off and get  worse with constantly changing workloads and systems in cloud environments. Big data systems, like \scope{}, further suffer from the widespread use of custom user code that ends up as black boxes in the cost models. 

One could consider improving a cost model by considering 
newer hardware and software configurations, such as machine SKUs, operator implementations, or workload characteristics. \scope{} team did attempt this path and put in significant efforts to improve their default cost model. 
This alternate cost model is available for \scope{} queries under a flag.
We turned this flag on and compared the costs from the improved model with the default one. Figure~\ref{fig:costModelwithCardFeedback}b 
shows the alternate model in solid blue line.
We  see that the correlation improves from $0.04$ to $0.10$ and the ratio curve for the manually improved cost model shifts a bit up, i.e., it reduces the over-estimation. However, it still suffers from the wide gap between the estimated and actual costs, again indicating that cost modeling is non-trivial in these environments. 

Finally, as discussed in the introduction and shown as dotted lines in Figure~\ref{fig:costModelwithCardFeedback}(b), fixing cardinalities
\rev{to the perfect values, i.e., that best that any cardinality estimator~\cite{stillger2001leo, cardLearner, lightweightCardModels, corrJoinsCardModels} could achieve,}
does not fill the gap between the estimated and the actual costs in \scope{}-like systems.

\vspace{-5pt}
\vspace{-3pt}
\section{Learned Cost Models}
\label{sec:learnedcosts}
\rev{In this section, we describe how we can leverage the common sub-expressions abundant in big data systems to learn a large set of smaller-sized but highly accurate cost models.}

\techreport{We note that it is practically infeasible to learn a single global model that is equally effective for all operators. This is why even traditional query optimizers model each operator separately.A single model is prone to errors because operators can have very different performance behavior (e.g., hash group by versus merge join), and even the performance of same operator can vary drastically depending on interactions with underneath operators via pipelining, sharing of sorting and grouping properties, as well as the underlying software or hardware platform (or the cloud instance). In addition, because of the complexity, learning a single model requires a large number of features, that can be prohibitively expensive to extract and combine for every candidate operator during query optimization.}

\vspace{-8pt}
\subsection{Specialized Cost Models}

\shep{As described in Section~\ref{sec:recurring}, shared cloud environments often have a large portion of recurring analytical queries (or jobs),
i.e.,  the same business logic is applied to newer instances of the datasets that arrive at regular intervals (e.g., daily or weekly).
Due to shared inputs, such recurring jobs often end up having one or more common subexpressions across them.}
For instance, 
the SCOPE query processing system at Microsoft has 
more than $50\%$ of jobs as recurring, with a large fraction of them appearing daily~\cite{jyothi2016morpheus}, and as high as 60\% having common subexpressions between them~\cite{cloudviews,bigsubs,cardLearner}.
Common subexpression patterns have also been reported in other production workloads, including 
Spark SQL queries in Microsoft's HDInsight~\cite{sparkcruise}, 
SQL queries from risk control department at Ant Financial Services Group~\cite{equitas}, and
iterative machine learning workflows~\cite{helix}.

\begin{figure}
	\centerline {
		\hbox{\resizebox{0.58\columnwidth}{!}{\includegraphics{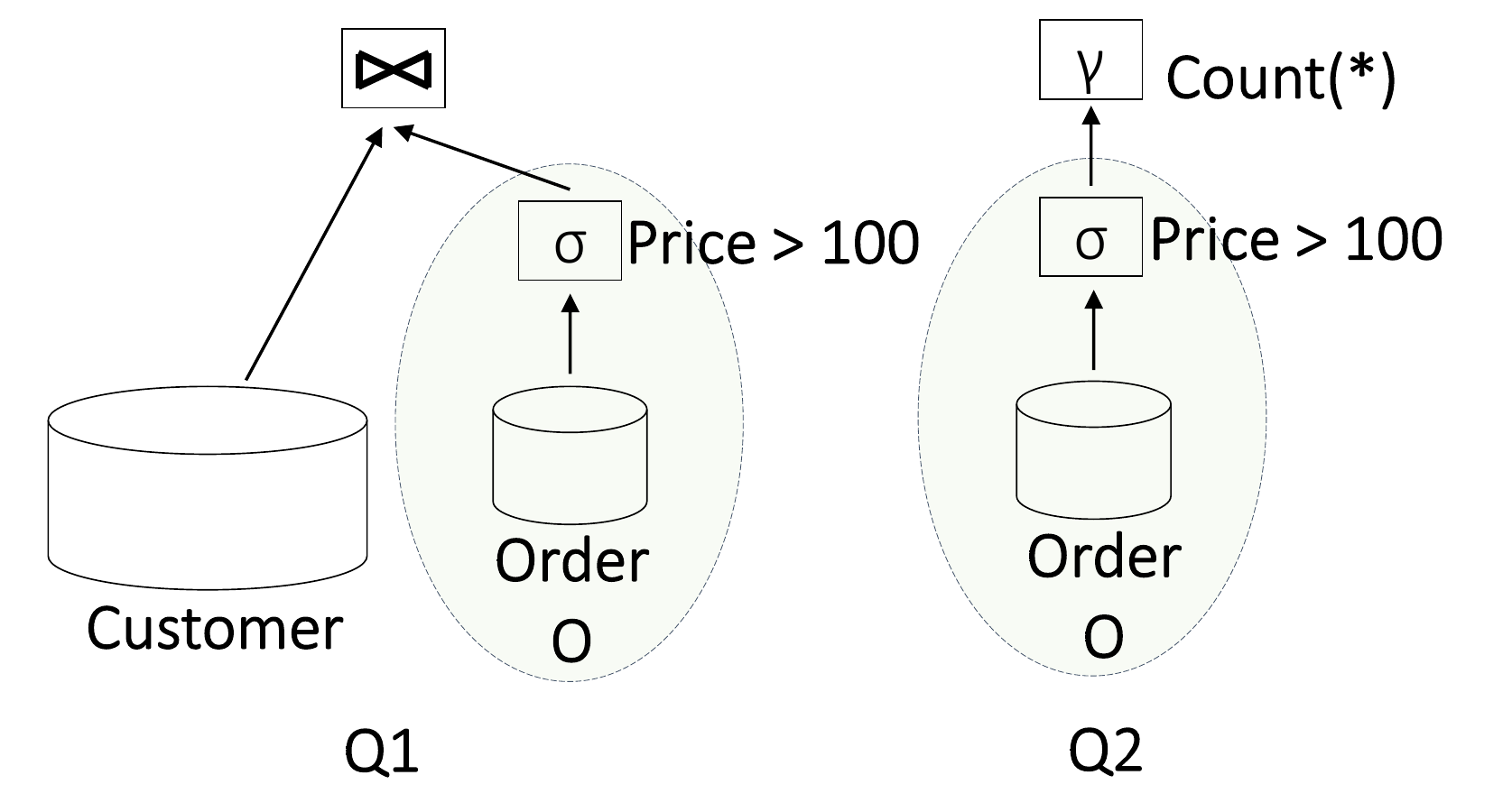}}}}
	\vspace{-10pt}
	\caption{Illustrating common subexpressions.}
	\vspace{-5.pt}
	\label{fig:commonSubs}
	\vspace{-8pt}
\end{figure} 

Figure~\ref{fig:commonSubs} illustrates a common subexpression, consisting of a scan followed by a filter, between two queries.
We exploit these common subexpressions by learning a large number of  {\it specialized} models, one for each unique {\it operator-sub- graph} template representing the subexpression. An  operator-subgraph template covers the root physical operator (e.g., Filter) and all prior (descendants) operators (e.g., scan) from the root operator of the subexpression. 
\shep{However, parameters and inputs in operator-subgraphs \emph{can vary over time}, and are used as features for the model (along with other logical and physical features) as discussed in Section~\ref{sec:featuresel}.} 

The operator-subgraph templates essentially \emph{capture the context of root operator, i.e, learn the behavior of root physical operator conditioned on the operators beneath it in the query plan}. This is helpful because of two reasons. First, the execution time of an operator depends on whether it is running in a pipelined manner, or is blocked until the completion of underneath operators. For example, the latency of a hash operator running on top of a filter operator is typically smaller compared to when running over a sort operator.  Similarly, the grouping and sorting properties of operators beneath the root operator can influence the latency of root operator~\cite{cascades95}.

Second, the estimation errors (e.g., of cardinality) grow quickly as we move up the query plan, with each intermediate operator building upon the errors of  children operators.  
The operator-subgraph models mitigates this issue partially since the intermediate operators are fixed and the cost of root operator depends only on the leaf level inputs. 
Moreover, when the estimation errors  are systematically off by certain factors (e.g., 10x), the subgraph models can adjust the weights such that the predictions are close to actual (discussed subsequently in Section~\ref{sec:learningmodel}). This is similar to adjustments learned explicitly in prior cardinality estimation work~\cite{stillger2001leo}. 
These adjustments generalize well since recurring jobs share similar schemas and the data distributions remain relatively stable, even as the input sizes change over time. \rev{Accurate cardinality estimations are, however, still needed in cases where simple adjustment factors do not exist~\cite{cardLearner}.}

Next, we discuss the learning settings, feature selection, and our choice of learning algorithm for operator-subgraphs.
\techreport{In Section~\ref{sec:opt-intg}, we describe the training and integration of learned models with the query optimizer.}

\vspace{-5pt}
\subsection{Learning Settings}

\noindent\textit{Target variable.} Given an operator-subgraph template, we learn the \emph{exclusive cost} of the root operator as our target. 
At every intermediate operator, we predict the exclusive cost of the operator conditioned on the subgraph below it. The exclusive cost is then combined with the costs of the children subgraphs to compute the total cost of the sub-graph, similar to how default cost models combine the costs.

\eat{
Another option is to directly learn the total cost of the subgraph, however, learning the total cost is less accurate because we often need to vary and select the most optimal physical properties of root operator (e.g., partition counts / degree of parallelism) that only change the behavior of root operator, while the underneath operators remain unchanged. In a top-down cascade-type query optimization, we estimate the cost of a plan in a bottom-up fashion  starting from leaf operators ~\cite{cascades95}. Thus, for an intermediate operator, we first predict the exclusive cost of the operator conditioned on  already chosen underneath operators, then add it to  the maximum of the total estimated costs of the children subgraphs.  This is quite similar to how default cost models combine the costs.
}

\begin{table}
	\large
	\resizebox{0.50\columnwidth}{!}{%
		\begin{tabular}{ l|l } 
			\hline
			Loss Function & Median Error\\ \hline 
			Median Absolute Error & 246\% \\ \hline 
			Mean Absolute Error & 62\% \\ \hline 
			Mean Squared Error & 36\%	\\ \hline 
			Mean Squared-Log Error  & 14\% \\ \hline 
		\end{tabular}%
	}
	\caption{\small Median error using 5-fold CV over the production workload for regression loss functions}
	\label{tab:loss_functions}
	\vspace{-25pt}
\end{table}

\textit{Loss function.} 
As the loss function, we use mean-squared log error between the predicted exclusive cost ($p$) and actual exclusive latency ($a$) of the operator:  $\frac{\sum_n(log(p+1)-log(a+1))^2}{n}$, here $1$ is added for mathematical convenience. 
Table~\ref{tab:loss_functions} compares the average median errors using 5-fold cross validation (CV)  of mean-squared log error with other commonly used regression loss functions, using elastic net as the learning model (described subsequently Section~\ref{sec:learningmodel}). We note that not taking the log transformation makes learning more sensitive to extremely large differences between actual and predicted costs. However, large differences often occur when the job's running time itself is long or even due to outlier samples because of machine or network failures (typical in big data systems). We, therefore, minimize the relative error (since $log(p+1)-log(a+1) = log(\frac{p+1}{a+1})$), that reduces the penalty for large differences. Moreover, our chosen loss function helps penalize under-estimation more than over-estimation, since under estimation can lead to under allocation of resources which is typically decided based on cost estimates.  Finally, log transformation implicitly ensures that the predicted costs are always positive.

\vspace{-5pt}
\subsection{Feature Selection}
\label{sec:featuresel}

It is expensive to extract and combine a large number of features every time we predict the cost of an operator --- a typical query plan in big data systems can involve 10s of physical operators, each of which can have multiple possible candidates. Moreover, large feature sets require more number of training samples, while many operator-subgraph instances typically have much fewer samples. Thus, we perform an offline analysis to identify a small set of useful features.

For selecting features, we start with a set of basic statistics that are frequently used for estimating costs of operators in the default cost model. These include the cardinality, the average row length, and the number of partitions.  We consider three kinds of cardinalities: 1) base cardinality: the total input cardinality of the leaf operators in the subgraph, 2) input cardinality: the total input cardinality from the children operators, and 3) output cardinality: the output cardinality of the operator-subgraph. We also consider normalized inputs (ignoring dates and numbers) and parameters to the job that typically vary over time for the recurring jobs. 
\techreport{We ignore features such as job name or cluster name, since they could induce strong bias and make the model brittle to the smallest change in names.}  

We further combine basic features to create additional derived features to capture the behavior of operator implementations and other heuristics used in default cost models.  We start a large space of possible derivations by applying (i) logarithms, square root, squares, and cubes of basic features, (ii) pairwise products among basic features and derived features listed in (i), and (iii)  cardinality features divided by the number of partitions (i.e. machine).  Given this set of candidate features, we use a variant of elastic net~\cite{zou2005regularization} model to select a subset of useful features that have at least one non-zero weight over all subgraph models. 


\begin{table}
	\resizebox{\columnwidth}{!}{%
		\begin{tabular}{ l|l } 
			\hline
			Feature & Description \\ \hline
			Input Cardinality (I) & Total Input Cardinality from children operators  \\ \hline
			Base Cardinality (B) & Total Input Cardinality at the leaf operators   \\ \hline
			Output Cardinality (C) & Output Cardinality from the current operator  \\ \hline
			AverageRowLength (L) & Length (in bytes) of each tuple  \\ \hline
			Number of Partitions (P) & Number of partitions allocated to the operator  \\ \hline
			Input (IN) & Normalized Inputs (ignored dates, numbers) \\ \hline
			Parameters (PM) & Parameters \\ \hline
		\end{tabular}%
	}
	\caption{\small Basic Features}
	\label{tab:basicfeatures}
	\vspace{-22.5pt}
\end{table}

\begin{table}
	\resizebox{\columnwidth}{!}{%
		\begin{tabular}{ p{3cm}|p{8cm} } 
			\hline
			Category & Features \\ \hline 
			Input or Output data & $\sqrt I$, $\sqrt B$, L*I, L*B, L*log(B),  L*log(I), L*log(C)		\\ \hline
			Input $\times$ Output & B*C,I*C,B*log(C),I*log(C),log(I)*log(C),I*C, log(B)*log(C)  \\ \hline 
			Input or Output per partition&  I/P, C/P,  I*L/P, C*L/P, $\sqrt I$/P,  $\sqrt C$/P,log(I)/P \\ \hline 
		\end{tabular}%
	}
	\caption{\small Derived Features }
	\vspace{-12.5pt}
	\label{tab:derivedfeatures}
	\papertext{\vspace{-15pt}}
	\techreport{\vspace{-15pt}}
\end{table}

Table~\ref{tab:basicfeatures} and Table~\ref{tab:derivedfeatures} depict the selected basic and derived features with non-zero weights.
We  group the derived features into three categories (i) input or output data, capturing the amount of data read, or written, 
(ii)  the product of input and output, covering the data processing and network communication aspects, and finally (iii) per-machine input or output, capturing the partition size. 

\begin{figure}
	\centerline {
		\hbox{\resizebox{0.97\columnwidth}{!}{\includegraphics{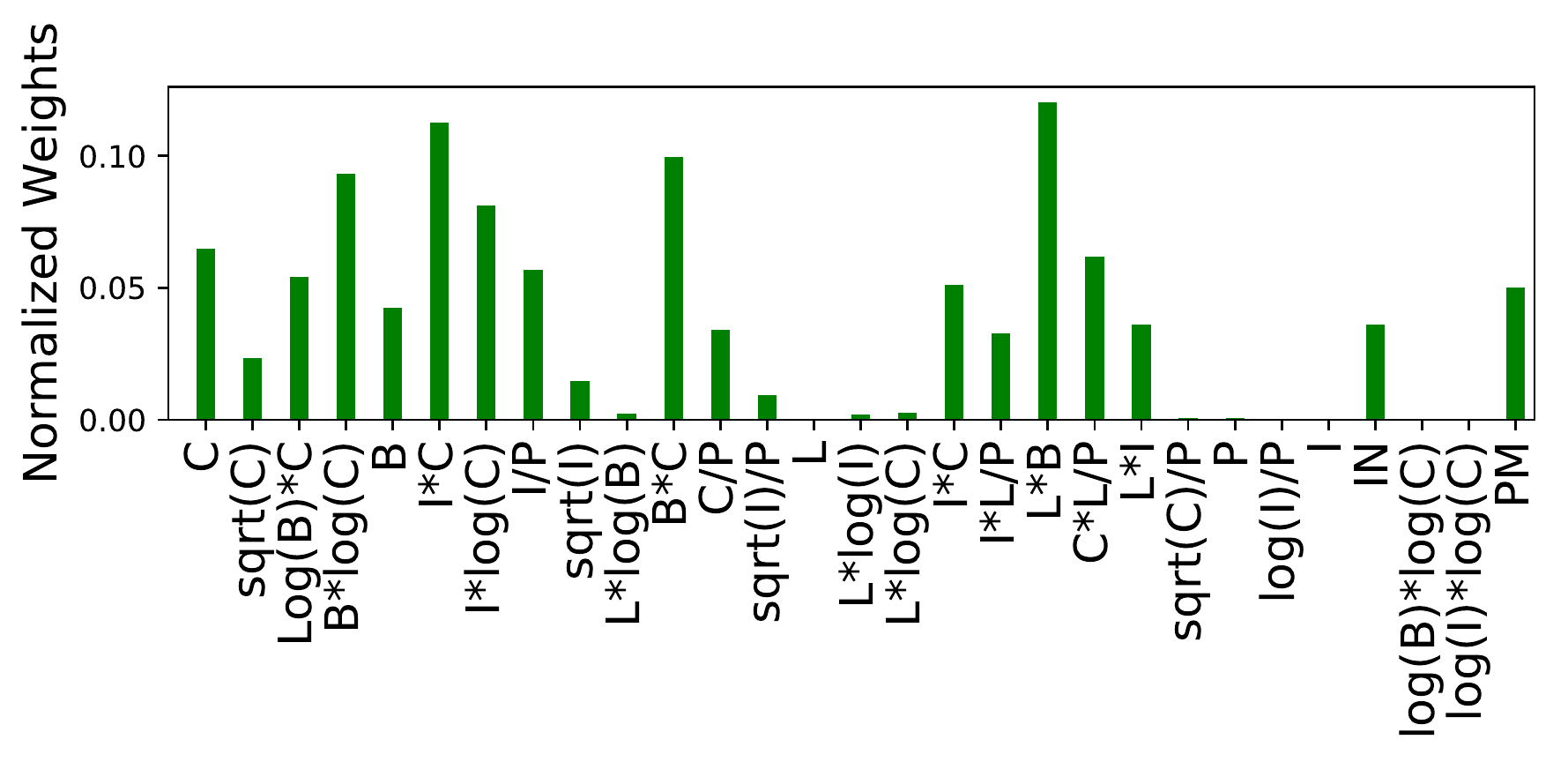}}}}
	\vspace{-16pt}	
	\caption{\small Features weights (Op-Subgraph model)}
	\label{fig:peroperatorsubgraph}
	\papertext{\vspace{-20pt}}	
\end{figure}

Further, we analyze the influence of each feature. While the influence of each feature varies over different subgraph models, Figure~\ref{fig:peroperatorsubgraph} shows the aggregated influence over all subgraph models of each feature. Given a total of $K$ non zero features and $N$ subgraph models, with $w_{in}$ as the weight of feature $i$ in model $n$, we measure the influence of feature $i$ using normalized weight $nw_i$, as $nw_i =  \frac{\sum_N abs(w_{in})}{\sum_{K}\sum_{N} abs(w_{kn})}$.

\eat{
In Table~\ref{tab:derivedfeatures}, we  group the derived features into three categories.  The first category has features where cardinalities are multiplied with the row length to capture the amount of data read, written, or transferred over the cluster. The second category consists of the product of input or base cardinalities with the output cardinalities, together covering the data processing and network communication aspects. The third category is unique to parallel big data systems capturing the per-machine input and output cardinality, as well as per machine input and output partition size.
}

\vspace{-8pt}
\subsection{Choice of learning model}
\label{sec:learningmodel}  
For learning costs over operator-subgraphs, we considered a number of variants of  linear-regression, SVM, decision tree and their ensembles, as well as neural network models. On 5-fold  cross-validation over our production workload, the following models give more accurate results compared to the default cost model:
(i) Neural network. 3-layers, hidden layer size = 30, solver = adam, activation = relu, l2 regularization = 0.005,
(ii) Decision tree: depth =15,
(ii) Random forest number of trees = 20, depth = 5,
(iii) FastTree Regression (a variant of Gradient Boosting Tree): number of trees = 20, depth = 5, and
(iv) Elastic net: $\alpha$ =1.0, fit intercept=True, l1 ratio=0.5.

We observe that the strict structure of subgraph template helps reduce the complexity, making the simpler models, e.g., linear- and decision tree-based regression models,  perform reasonably well with the chosen set of features. A large number of  operator-subgraph templates have  fewer training samples,  e.g., more than half of the subgraph instances have $< 30$ training samples for the workload described in Section~\ref{sec:motivation}. In addition, because of the variance in cloud environments (e.g., workload fluctuations, machine failures, etc.), training samples  can have noise both in their features (e.g., inaccurate statistics estimates) and the class labels (i.e., execution times of past queries). Together, both these factors lead to over-fitting, making complex models such as neural network as well as ensemble-based models such as gradient-boost perform worse.

\textbf{Elastic net}~\cite{zou2005regularization}, a $L_1$ and $L_2$ regularized linear regression model, on the other hand, is relatively less prone to overfitting. In many cases, the number of candidate features (ranging between $25$ to $30$) is as many as the number of samples, while only  a select few features are usually relevant for a given subgraph. The relevant features further tend to differ across subgraph instances. Elastic net helps perform \emph{automatic feature selection}, by selecting a few relevant predictors for each subgraph independently. Thus, we train all subgraphs with the same set of features, and let elastic net select the relevant ones. Another advantage of elastic net model is that it is intuitive and easily interpretable, like the default cost models which are also  weighted sums of a number of statistics. This is an important requirement for  effective debugging and analysis of production jobs.

\rev{Table~\ref{tab:mlmodelscv} depicts the Pearson correlation and median error  of the five machine learning models over the production workload. We see that operator-subgraphs models trained using elastic net can make sufficiently accurate cost predictions (14\% median error), with a high correlation (more than $0.92$) with the actual runtime, a substantial improvement over the default cost model (median error of $258$\% and Pearson correlation of $0.04$). In addition, elastic net models are fast to invoke during query optimization, and have low storage footprint that indeed makes it feasible for us to learn specialized models for each possible subgraph.}

\vspace{-5pt}
\section{Robustness}
\label{sec:robustness}
We now discuss how we learn {\it robust cost models}. 
As defined in Section~\ref{sec:intro}, robust cost models cover the entire workload with high accuracy  for a substantial time period before requiring retraining.
\techreport{
In contrast to prior work on robust query processing \cite{markl2004robust, dutt2016plan, lookahead-ip} that either modify the plan during query execution, or execute multiple plans simultaneously, we leverage the massive cloud workloads to learn robust models offline and integrate them with the optimizer to generate one robust plan with minimum runtime overhead. In this section, we first explain the coverage and accuracy tradeoff for the operator-subgraph model.
}
\papertext{ We first explain the coverage and accuracy tradeoff for the operator-subgraph model.}
Then, we discuss the other extreme, namely an {\it operator} model, and introduce additional models to bridge the gap between the two extremes. Finally, we discuss how we combine predictions from individual models to achieve robustness. 

\eat{
\begin{figure*}
		\vspace{-33pt}
		\centerline {
			\hbox{\resizebox{0.80\textwidth}{!}{\includegraphics{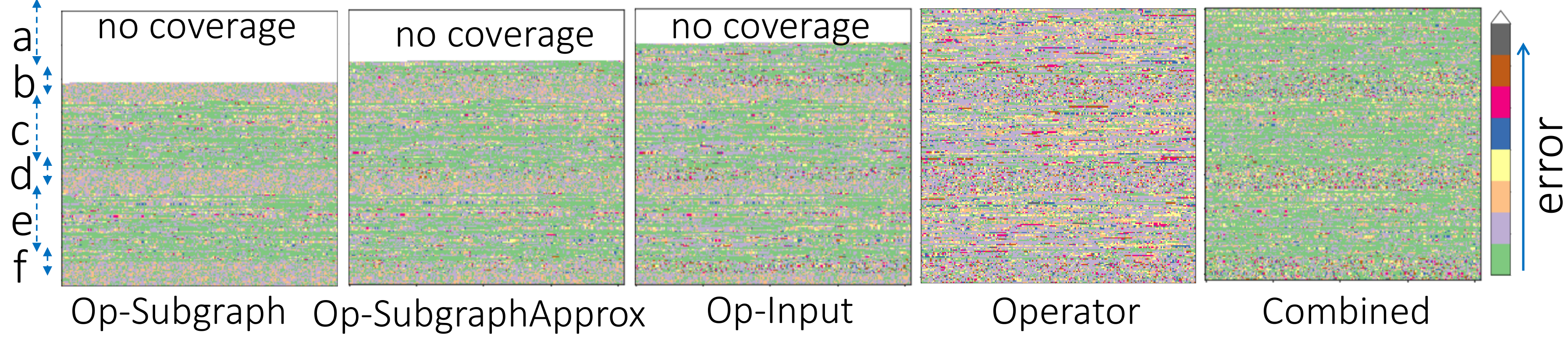}}}}
	   \vspace{-10pt}
		\caption{{\small Heatmap of errors over $42K$ operators from production jobs. Each point in the heatmap depicts the error of learned cost with respect to the actual runtime}}
		\vspace{-14pt}
		\label{fig:cov_accuracy_heatmap}
\end{figure*}
}

\vspace{-0.4cm}
\subsection{Accuracy-Coverage Tradeoff}
\label{sec:tradeoff}

The operator-subgraph model presented in Section~\ref{sec:learnedcosts} is highly specialized. As a result, this model is likely to be highly accurate. Unfortunately, the operator-subgraph model does not cover subgraphs that are not repeated in the training dataset,
i.e., it has limited coverage. \rev{For example, over $1$ day of \microsoft{} production workloads,  operator-subgraphs have learned  models for only 54\% of the subgraphs. Note that we create a learned model for a subgraph if it has at least $5$ occurrences over the single day worth of training data.} Thus, it is difficult to predict costs for arbitrary query plans consisting of subgraphs never seen in training dataset.

\stitle{The other extreme: Operator model.}  In contrast to the operator-subgraph model, we can learn a model for each physical operator, similar to how traditional query optimizers model the cost. \shep{The operator models estimate the execution latency of a query by composing the costs of individual operators in a hierarchical manner akin to how default cost models derive query costs. As a result, operator models can  predict the cost of any query in the workload, including those previously unseen in the training dataset.} However, similar to traditional cost model, operator models also suffer from poor accuracy 
since the behavior of an operator changes based on what operations appear below it. Furthermore, the estimation errors in the features or statistics at the lower level operators of a query plan are propagated to the upper level operators that significantly degrades the final prediction accuracy. \shep{On $5$-fold cross-validation over $1$ day of \microsoft{} production workloads, operator models results in $42\%$ median error and $0.77$ Pearson correlation, which although better than the default cost model (258\% median error and 0.04 Pearson correlation), is relatively lower compared to that of operator-subgraph models (14\% median error and $0.92$ Pearson correlation).}
Thus, there is an accuracy and coverage tradeoff when learning cost models, with operator-subgraph and operator models being the two extremes of this tradeoff.

\vspace{-0.4cm}
\subsection{Bridging the Gap}

We now present additional models that fall between the two extreme models in terms of the accuracy-coverage trade-off.

\vspace{0.1cm}
\noindent{\bf Operator-input model.}  An improvement to per-operator is to learn a model for all jobs that share similar inputs. Similar inputs also tend to have similar schema and similar data distribution even as the size of the data changes over time, thus operator models learned over similar inputs often generalize over future job invocations.
In particular, we create a model for each operator and input template combination. An input template is a normalized input where we ignore the dates, numbers, and parts of names that change over time for the same recurring input, thereby allowing grouping of jobs that run on the same input schema over different sessions. Further, to partially capture the context, we featurize the intermediate subgraph by introducing two additional features: 1) the number of logical operator in the subgraph (CL) and 2) the depth  of the physical operator in the sub-graph (D). This helps in distinguishing subgraph instances that are extremely different from each other.

\begin{table}
\papertext{\vspace{-5pt}}
	\begin{center}
		\resizebox{0.65\columnwidth}{!}{%
			\begin{tabular}{ p{3cm}|l|l} 
				\hline
				Model &  Correlation  &  Median Error\\ \hline 
				Default  & 0.04 & 258\% \\ \hline
				Neural Network  & 0.89 & 27\% \\ \hline
				Decision Tree & 0.91 & 19 \%\\ \hline
				Fast-Tree regression & 0.90 & 20\% \\ \hline
				Random Forest &  0.89 & 32\% \\  \hline
				\textbf{Elastic net} & \textbf{0.92} & 14\% \\ \hline
			\end{tabular}%
		}
		\caption{Correlation and error w.r.t. actual runtime for the operator-subgraphs}
		\vspace{-30pt}
		\label{tab:mlmodelscv}
	\end{center}
\end{table}

\techreport{

\begin{figure*}
	\vspace{-20pt}
	\centerline {
		\hbox{\resizebox{0.7\textwidth}{!}{\includegraphics{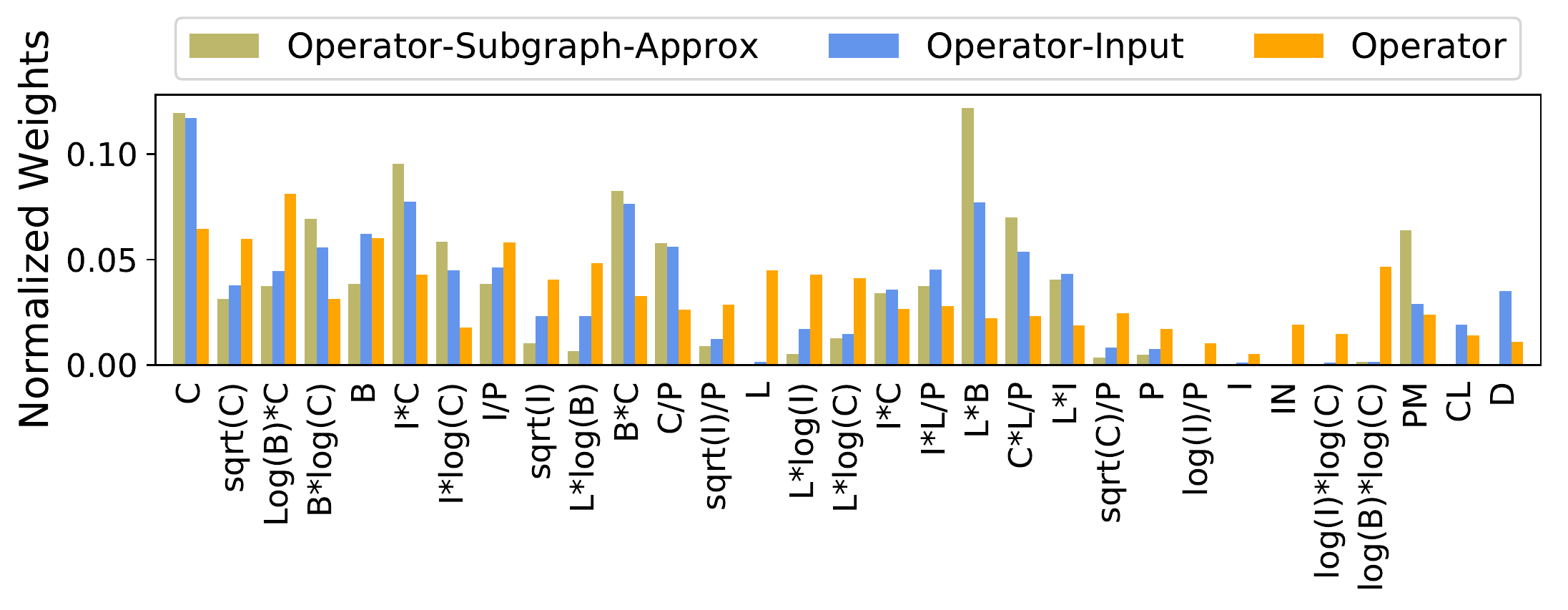}}}}	
		\vspace{-32pt}
	\caption{Feature weights (all other models)}
	\label{fig:mergedfeatures_imp}
\end{figure*}
}

\papertext{
\begin{figure*}
	\hspace{-0.5cm}
	\begin{minipage}{0.65\textwidth}
		\vspace{-20.5pt}
		\centerline {
			\hbox{\resizebox{\columnwidth}{!}{\includegraphics{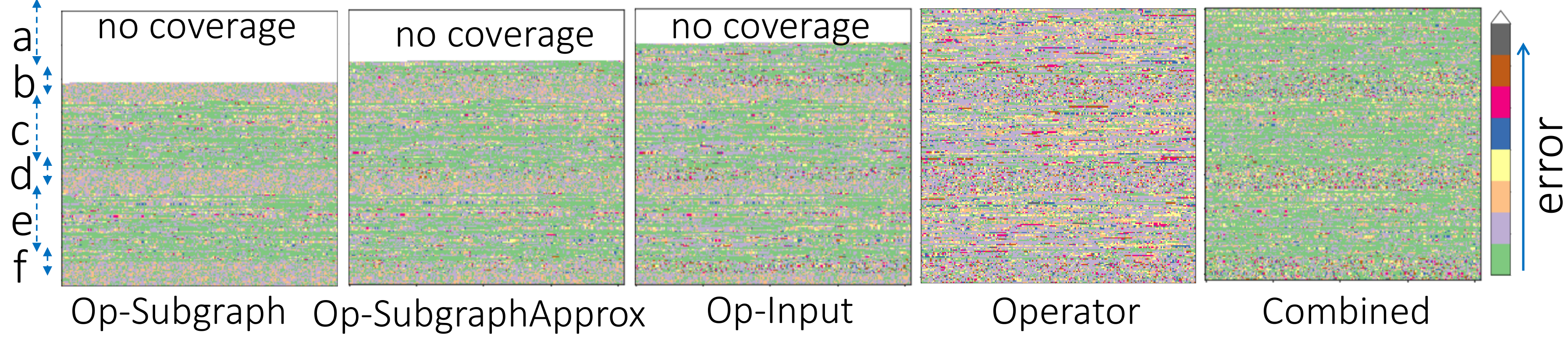}}}}
		\vspace{-10pt}
		\caption{\small Heatmap of errors over $42K$ operators from production jobs. Each point in the heatmap depicts the error of learned cost with respect to the actual runtime}
		\vspace{-25pt}
		\label{fig:cov_accuracy_heatmap}
	\end{minipage}
	\hspace{0.15cm}
	\begin{minipage}{0.35\textwidth}
		\vspace{-1pt}
		\begin{center}
			\small
			\resizebox{\columnwidth}{!}{%
				\begin{tabular}{ p{2cm}|p{1.5cm}|p{1.2cm}|l} 
					\hline
					Model & Correlation &  Median Error & Coverage \\ \hline 
					\textbf{Default}  & \textbf{0.04} & \textbf{258\%} &  \textbf{100\%} \\ \hline
					Op-Subgraph  & 0.92 & 14\% & 54\% \\ \hline
					Op-Subgraph Approx & 0.89 & 16 \% & 76\%\\ \hline
					Op-Input & 0.85 & 18\% & 83\% \\ \hline
					Operator & \textbf{0.77} & \textbf{42\%} & \textbf{100\%} \\ \hline
					\textbf{Combined} &  \textbf{0.84} & \textbf{19\%}  & \textbf{100\%} \\  \hline
				\end{tabular}%
			}
			\captionof{table}{\small Performance of learned models w.r.t to actual runtimes}
			\label{tab:modelaccuracies}
			\vspace{-25pt}
		\end{center}	
	\end{minipage}
\end{figure*}
}

\vspace{0.1cm}
\noindent{\bf Operator-subgraphApprox model.} While operator-sub- graph models exploit the overlapping and recurring nature of big data analytics, there is also a large number (about 15-20\%) of  subgraphs that are similar but are not exactly the same. 
To bridge this gap, we relax the notion of subgraph similarity, and learn one model for all subgraphs that have the same inputs and the same \emph{approximate} underlying subgraph.  We consider two subgraphs to be approximately same if they have the same physical operator at the root, and consist of the \emph{same frequency of each logical operator in the underneath subgraph (ignoring the ordering between operators).}
\techreport{Thus, there are two relaxations: (1) we use frequency of logical operators instead of physical operators (note that this is one of the additional features in Operator-input model)  and  (ii)  we ignore the ordering between operators.} This relaxed similarity criteria allows grouping  of similar subgraphs without substantially reducing the coverage.  Overall, operator-subgraphApprox model is a hybrid of the operator-subgraph and operator-input models: it achieves much higher accuracy compared to operator or operator-input models, and more coverage compared to operator-subgraph model.

\vspace{0.1cm}
Table~\ref{tab:modelaccuracies} depicts the Pearson correlation, the median accuracy using 5-fold cross-validation, as well as the coverage of individual cost models using elastic net over production workloads.
As we move from more specialized to more generalized models (i.e., operator-subgraph to operator-subgraphApprox to operator-input to operator), we see that the model accuracy decreases while the coverage over the workload increases. 
\papertext{We provide more details on the importance of features for each model in our technical report~\cite{techreport}.}
\techreport{Figure~\ref{fig:mergedfeatures_imp} shows the feature weights for each of the intermediate models.   We see that while the weight for specialized models, like the operator-subgraph model (Figure~\ref{fig:peroperatorsubgraph}), are concentrated on a few features, the  weights for more generalized models, like the per-operator model, are more evenly distributed.}

\vspace{-0.2cm}
\subsection{The Combined Model}
\label{sec:combined}

Given multiple learned models, with varying accuracy and coverage, the strawman approach is to select a learned model in decreasing order of their accuracy over training data, starting from operator-subgraphs to operator-subgraphApprox to operator-inputs  to operators. However, as discussed earlier, there are subgraph instances where more accurate models result in poor performance over test data due to over-fitting. 

\techreport{
	
	\begin{figure*}
		\centerline {
			\hbox{\resizebox{0.8\textwidth}{!}{\includegraphics{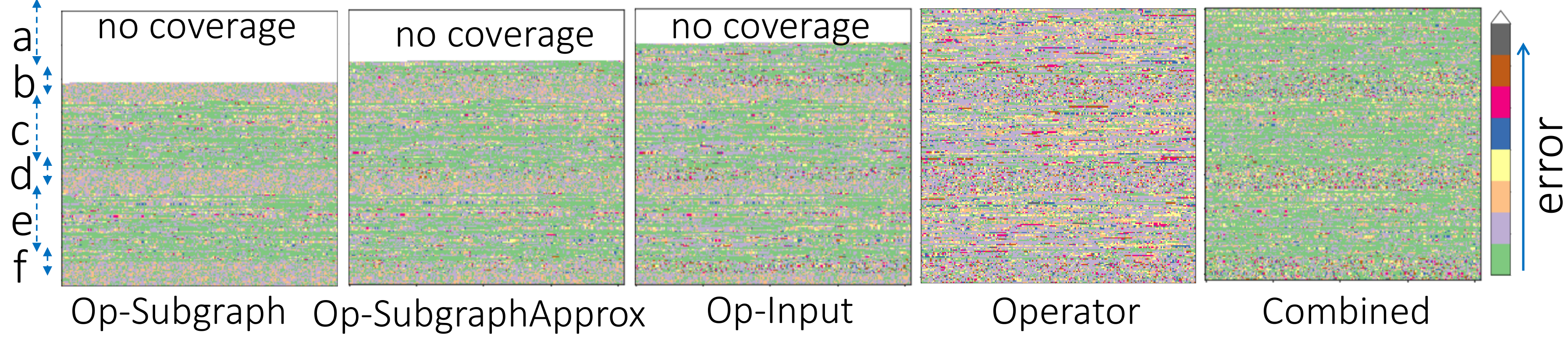}}}}
		\vspace{-10pt}
		\caption{ Heatmap of errors over $42K$ operators from production jobs. Each point in the heatmap depicts the error of learned cost with respect to the actual runtime}
		\label{fig:cov_accuracy_heatmap}
		\vspace{-10pt}
	\end{figure*}
	
	\begin{table}
		\begin{center}
			\small
			\resizebox{0.75\columnwidth}{!}{%
				\begin{tabular}{ p{2cm}|p{1.5cm}|p{1.2cm}|l} 
					\hline
					Model & Correlation &  Median Error & Coverage \\ \hline 
					\textbf{Default}  & \textbf{0.04} & \textbf{258\%} &  \textbf{100\%} \\ \hline
					Op-Subgraph  & 0.92 & 14\% & 54\% \\ \hline
					Op-Subgraph Approx & 0.89 & 16 \% & 76\%\\ \hline
					Op-Input & 0.85 & 18\% & 83\% \\ \hline
					Operator & \textbf{0.77} & \textbf{42\%} & \textbf{100\%} \\ \hline
					\textbf{Combined} &  \textbf{0.84} & \textbf{19\%}  & \textbf{100\%} \\  \hline
				\end{tabular}%
			}
			\captionof{table}{\small Performance of learned models w.r.t to actual runtimes}
			\label{tab:modelaccuracies}
			\vspace{-30pt}
		\end{center}
	\end{table}
}

\eat{
The simplest way to select a learned model for prediction is to adopt a rule-based approach that looks for the presence of learned models in decreasing order of accuracies over training data, starting from operator-subgraphs to operator-subgraphApprox to operator-inputs  to operators.
However, as discussed earlier, there are subgraph instances, especially with insufficient training data, where more accurate models result in poor performance over test data due to over-fitting.
Another option is use to a validation set, and select models based on threshold in terms of accuracy. However, it is cumbersome and practically infeasible to decide one specific threshold to switch from one model to another that works (i) equally well for all subgraphs, and (ii) over a long prediction window. 
Moreover, if a model $x$ has a lower overall training accuracy than model $y$, there may still be some class of queries where model $x$ performs better than $y$ and vice-versa. This could be because of factors such as missing one or more informative features, or noise in feature values or class labels. In reality, the feature space over which the different models perform well is difficult to separate using a threshold or a rule, nor is it feasible to keep track of sub-spaces where each model performs better.
}

To illustrate, Figure~\ref{fig:cov_accuracy_heatmap} depicts the heat-map representation of the accuracy and coverage of different individual models,
over more than $42K$ operator instances from our production workloads. Each point in the heat-map represents the error of the predicted cost with respect to the actual runtime: the more green the point,  the less the error. The empty region at the top of the chart depicts that the learned model does not cover those subgraph instances.
We can see that the  predictions are mostly accurate for operator-subgraph models, while Operator models have relatively more error (i.e, less green) than the operator-subgraph models.
Operator-input have more coverage with marginally more error (less green) compared to operator-subgraphs. However, for regions  b, d, and f, as marked on the left side of the figure, we notice that operator-input performs better (more blue and green) than operator-subgraph. This is because operators in those regions have much fewer training samples that results in over-fitting. Operator-input, on the other hand, has more training samples, and thus performs better. Thus, it is difficult to decide a rule or threshold that always select the best performing model for a given subgraph instance.

\eat{
\begin{figure*}
	\hspace{-0.4cm}
	\begin{minipage}{0.75\textwidth}
	\vspace{-35pt}
	\centerline {
		\hbox{\resizebox{\columnwidth}{!}{\includegraphics{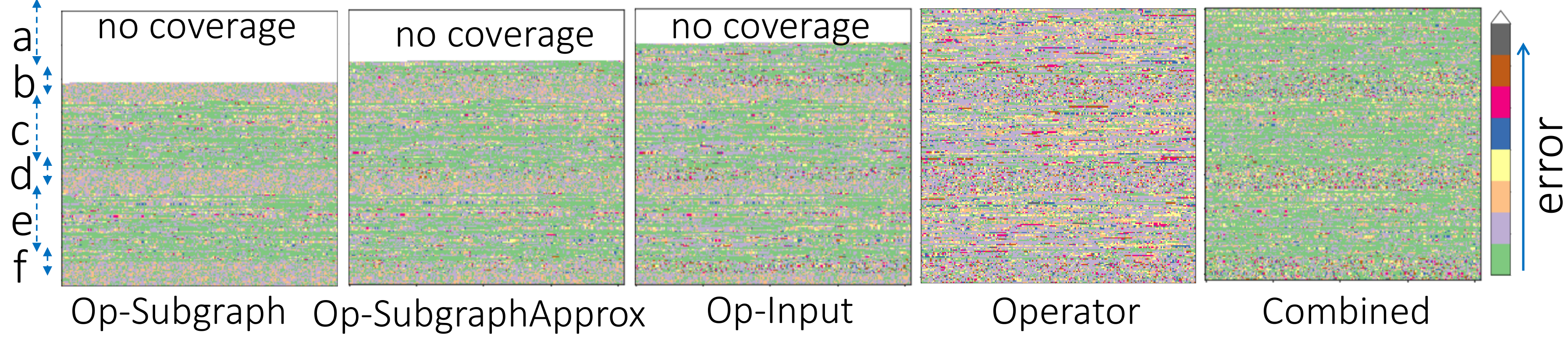}}}}
	\vspace{-8pt}
	\caption{{\small Heatmap of errors over $42K$ operators from production jobs. Each point\\ in the heatmap depicts the error of learned cost with respect to the actual runtime}}
	\vspace{-15pt}
	\label{fig:cov_accuracy_heatmap}
	\end{minipage}
	\begin{minipage}{0.25\textwidth}
	\begin{center}
		\resizebox{\columnwidth}{!}{%
			\begin{tabular}{ l|l|p{1cm}} 
				\hline
				Model &  Correlation  &  Median Error \\ \hline 
				Default  & 0.04 & 258\% \\ \hline
				Neural Network  & 0.79 & 31\% \\ \hline
				Decision Tree & 0.73 & 41 \%\\ \hline
				FastTree Regression & \textbf{0.84} & \textbf{19\%} \\ \hline
				Random Forest &  0.80 & 28\% \\  \hline
				\textbf{Elastic net} & 0.68 & 64\% \\ \hline
			\end{tabular}%
		}
		\captionof{table}{\small Correlation and accuracy relative to actual runtime latency for machine learning models used for the Combined Model}
		\vspace{-11.5pt}
		\label{tab:combinedmlcv}
	\end{center}
\end{minipage}

\end{figure*}
}

\eat{
\begin{figure}
	\centerline {
		\hbox{\resizebox{1.05\columnwidth}{!}{\includegraphics{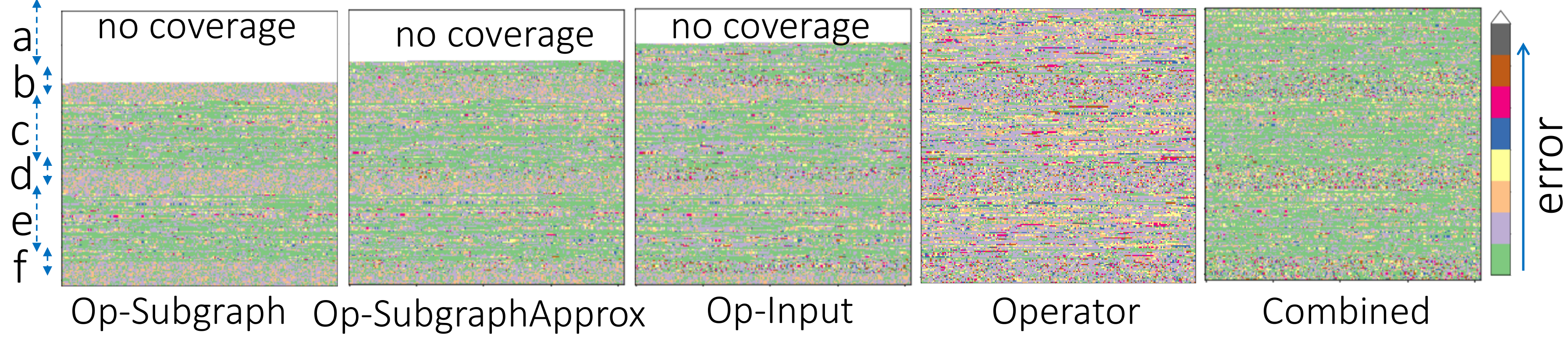}}}}
	\vspace{-8pt}
	\caption{{\small Heatmap of errors over $42K$ operators from production jobs. Each point in the heatmap depicts the error of learned cost with respect to the actual runtime}}
	\vspace{-15pt}
	\label{fig:cov_accuracy_heatmap}
\end{figure}
}

\begin{table}
	\begin{center}
		\resizebox{0.7\columnwidth}{!}{%
			\begin{tabular}{ l|l|p{2cm}} 
				\hline
				Model &  Correlation  &  Median Error \\ \hline 
				Default  & 0.04 & 258\% \\ \hline
				Neural Network  & 0.79 & 31\% \\ \hline
				Decision Tree & 0.73 & 41 \%\\ \hline
				FastTree Regression & \textbf{0.84} & \textbf{19\%} \\ \hline
				Random Forest &  0.80 & 28\% \\  \hline
				\textbf{Elastic net} & 0.68 & 64\% \\ \hline
			\end{tabular}%
		}
		\captionof{table}{\small Correlation and  errror w.r.t   actual runtimes for the Combined Model}
		\label{tab:combinedmlcv}
		\vspace{-30.5pt}
	\end{center}
\end{table}

\stitle{Learning a meta-ensemble model.} 
We introduce a meta-ensemble model that  uses the predictions from specialized models as meta features, along with the following extra features: (i) cardinalities (I, B, C), (ii) cardinalities per partition (I/P, B/P, C/P), and (iii) number of partitions (P) to  output a more accurate cost.
Table~\ref{tab:combinedmlcv} depicts the performance of different machine learning models that we use as a meta-learner on our production workloads. We see that the FastTree regression~\cite{fasttree}
results in the most accurate predictions.
FastTree regression is a variant of the gradient boosted regression trees ~\cite{friedman2002stochastic} that uses an efficient implementation of the MART gradient boosting algorithm ~\cite{mart}. It builds a series of regression trees (estimators), with each successive tree fitting on the residual of trees that precede it.  Using $5$-fold cross-validation, we find that the maximum of only $20$ regression trees with mean-squared log error as the loss function and the sub-sampling rate of $0.9$ is sufficient for optimal performance. 
As depicted in Figure~\ref{fig:cov_accuracy_heatmap},  using FastTree Regression as a meta-learner has three key advantages.

First,  FastTree regression can effectively characterize the space where each model performs well.
The regression trees  recursively split the space defined by the predictions from individual models and features, creating fine-grained partitions such that prediction in each partition are highly accurate. 


\eat{
\begin{table}
	\begin{center}
		\caption{Correlation and accuracy with respect to actual runtime latency for different machine learning models used for the Combined Model}
		\vspace{-11.5pt}
		\label{tab:combinedmlcv}
		\resizebox{0.85\columnwidth}{!}{%
			\begin{tabular}{ l|l|l} 
				\hline
				Model &  Pearson Correlation  &  Median Error (\%) \\ \hline 
				Default  & 0.04 & 258\% \\ \hline
				Neural Network  & 0.79 & 31\% \\ \hline
				Decision Tree & 0.73 & 41 \%\\ \hline
				Fast-Tree Regression & \textbf{0.84} & \textbf{19\%} \\ \hline
				Random Forest &  0.80 & 28\% \\  \hline
				\textbf{Elastic net} & 0.68 & 64\% \\ \hline
			\end{tabular}%
		}
	\end{center}
	\vspace{-10pt}
\end{table}
}

Second, FastTree regression performs better for operator instances where individual models perform worse.  For example, some of the red dots found in outlier regions b, d and f of the individual model heat-maps are missing in the combined model. This is because FastTree regression makes use of sub-sampling to build each of the regression trees in the ensemble, making it more resilient to overfitting and noise in execution times of prior queries. Similarly, for region a where subgraph predictions are missing, FastTree regression creates an extremely large number of fine-grained splits using extra features and operator predictions to give even better accuracy compared to Operator models.

Finally, \shep{the combined model is naturally capable of covering all possible plans, since it uses Operator model as one of the predictors.} Further, as depicted in Figure~\ref{fig:cov_accuracy_heatmap}, the combined model is comparable to that of specialized models, and almost always has better accuracy than Operator model. The combined model is also flexible enough to incorporate  additional models or features, or replace one model with another. However, on also including the default cost model, it did not result in any improvement on \scope{}. Next, we discuss how we integrate the learned models within the query optimizer.



\vspace{-10pt}
\section{Optimizer Integration}
\label{sec:opt-intg}
\vspace{-3pt}

In this section, we describe our two-fold integration of \system within \scope{}. First, we discuss our end-to-end feedback loop to learn cost models and generate predictions during optimization. Then, we discuss how we extend the optimizer for supporting resource-exploration using learned models.

\begin{figure*}
	\vspace{-5pt}
	\hspace{-0.9cm}
	\begin{subfigure}{0.265\textwidth}
		\centerline {
			\hbox{\resizebox{\columnwidth}{!}{\includegraphics{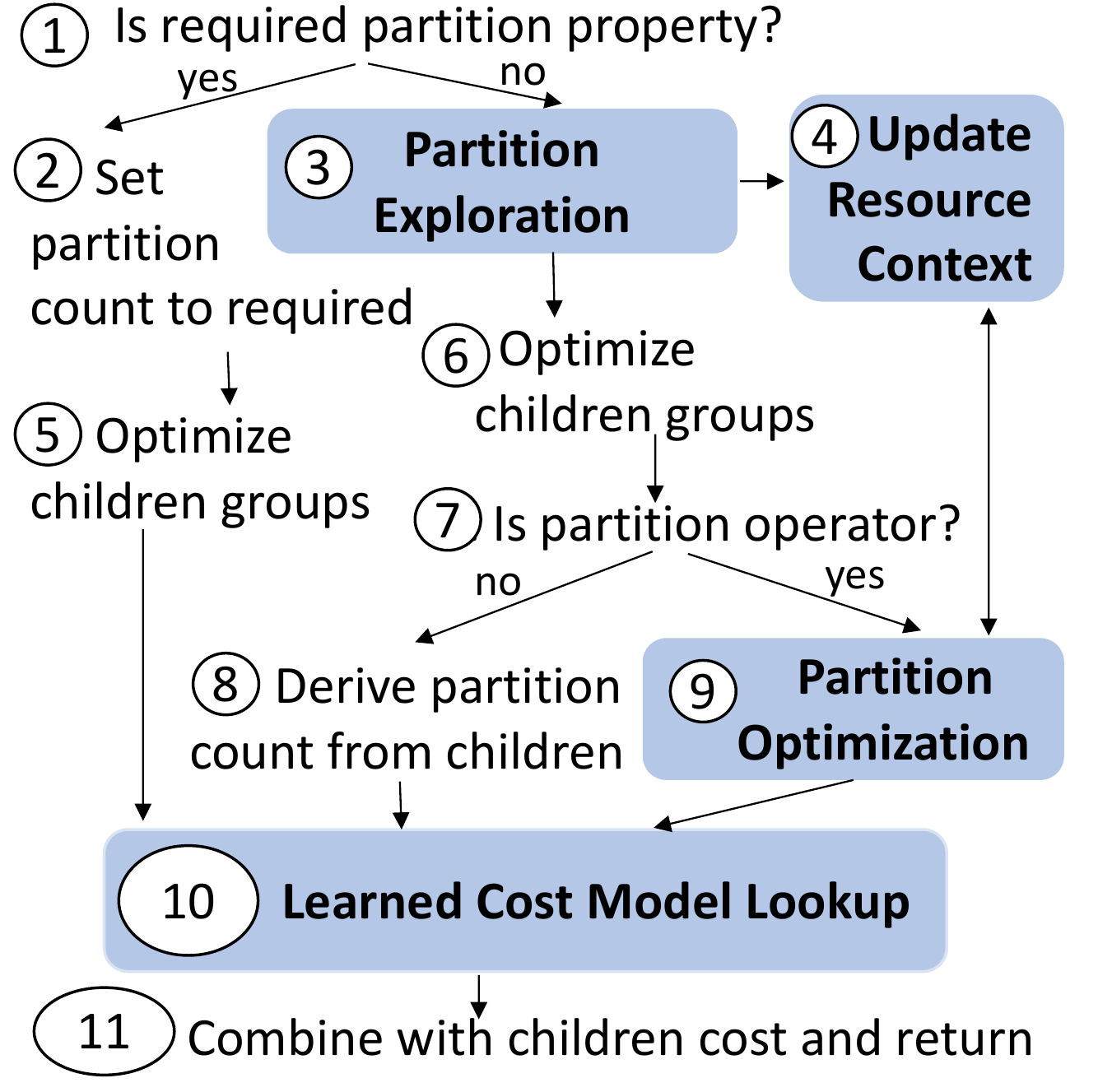}}}}
		\caption{Resource-aware planning}
		\label{fig:resplan}
	\end{subfigure}
	\hspace{-0.2cm}
	\begin{subfigure}{0.45\textwidth}
		\centerline {
			\hbox{\resizebox{\columnwidth}{!}{\includegraphics{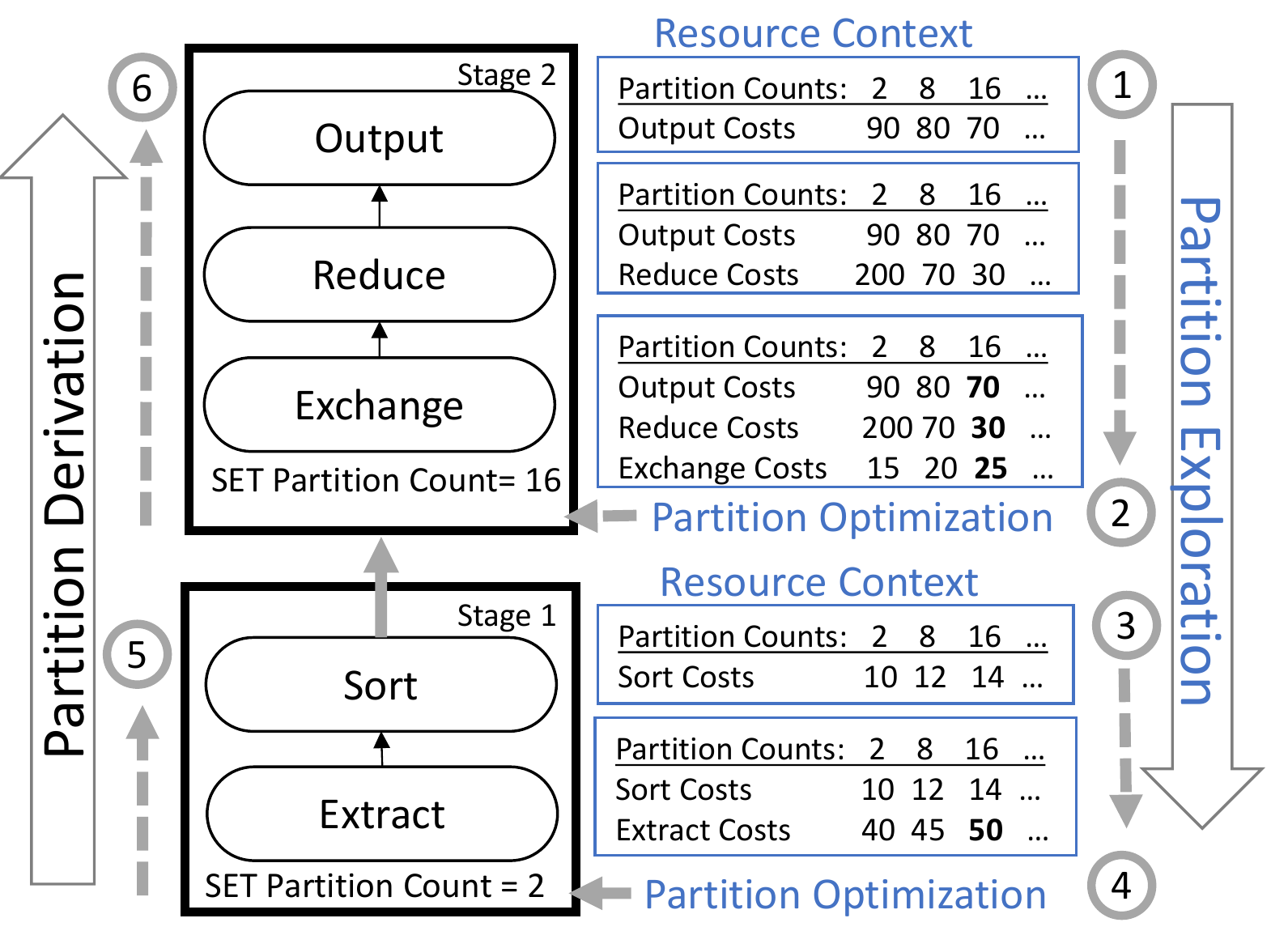}}}}
		\vspace{-5pt}
		\caption{Example query plan}
		\label{fig:resplan_example}	
	\end{subfigure}
	\hspace{-0.3cm}
	\begin{subfigure}{0.27\textwidth}
		\vspace{-18pt}
		\centerline {
			\hbox{\resizebox{\columnwidth}{!}{\includegraphics{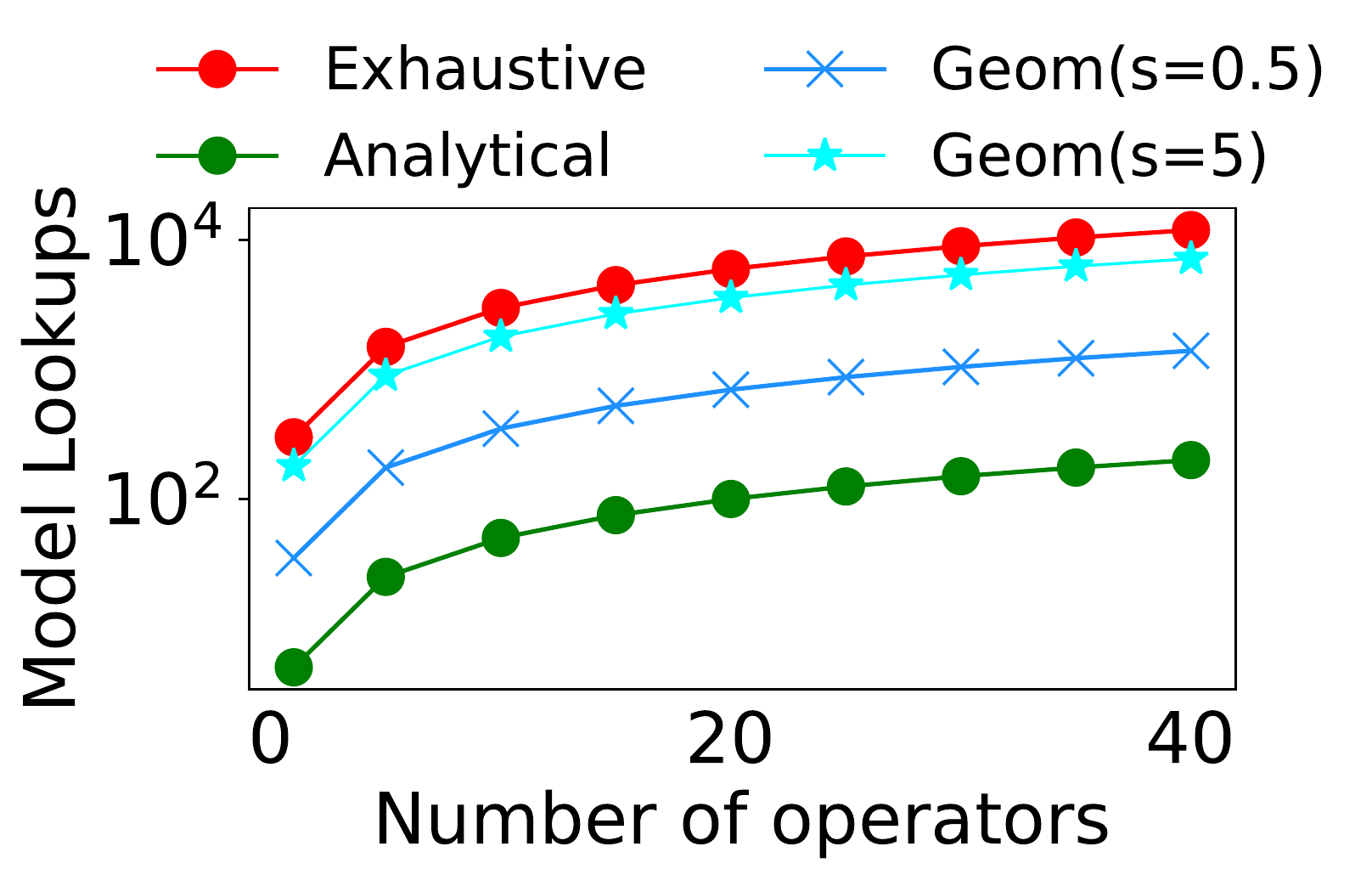}}}}
		\vspace{-8pt}
		\caption{Model look-ups for partition exploration}
		\label{fig:model_invocations}
	\end{subfigure}
	\vspace{-12pt}
	\caption{Integrating learned cost models with the query optimizer}
	\label{fig:learnedIntg}	
	\vspace{-10pt}
\end{figure*}

\vspace{-8pt}
\subsection{Integrating Learned Cost Models}
\label{trainfeedbak}
\rev{
The integration of learned models with the query optimizer involves the following three components.}

\vspace{0.05cm}
\noindent
\rev{
\textbf{Instrumentation and Logging.} Big data systems, such as \scope{}, are already instrumented to collect logs of query plan statistics such as cardinalities, estimated costs, as well as runtime traces for analysis and debugging. For uniquely identifying each operator and the subplan, query optimizers annotate each operator with a \emph{signature}~\cite{bruno2013continuous}, a 64-bit hash value that can be recursively computed in a bottom-up fashion by combining (i) the signatures of children operators, (ii) hash of current operator's name, and (iii) hash of operator's logical properties. 
We extend the optimizer to  compute three more signatures, 
one for each individual sub-graph mode, and extract additional statistics (i.e., features) that were missing for \system. Since all signatures can be computed simultaneously in the same recursion, and that there are only $25-30$ features (most of which were already extracted), the additional overhead is minimal ($\leq$ 10\%), as we describe in Section~\ref{sec:experiments}. This overhead includes both the logging and the model lookup that we discuss subsequently.
}

\vspace{0.02cm}
\noindent
\rev{
\textbf{Training and Feedback}.
Given the logs of past query runs, we learn each of the four individual elastic net-based models independently and in parallel using our \scope{}-based parallel model trainer. Experimentally, we found that a training window of two days and a training frequency of every ten days results in acceptable accuracy and coverage (Section~\ref{sec:experiments}). We then use the predictions from individual models on the next day  queries to learn the combined FastTree regression model. Since individual models can be trained in parallel, the training time is not much, e.g., it takes less than $45$ minutes for training $25K$ models from over $50,000$ jobs using a cluster of $200$ nodes.
Once trained, we serialize the models 
and feedback them to the optimizer. The models can be served either from a text file, using an additional compiler flag, or using a web service that is backed by a SQL database.}

\noindent
\rev{\textbf{Look-up}. All models relevant for a cluster are loaded upfront by the optimizer, into a hash map with keys as {\it signatures} of models, to avoid expensive lookup calls during optimization. When loaded simultaneously, all $25K$ models together take about $600$ MB of memory, which is within an acceptable range. \shep{Finally, for cost estimation, we modify the Optimize Input phase of Cascade optimizer to invoke learned models. Figure~\ref{fig:resplan} highlights the  the key steps that we added in blue.  Essentially,  we replace the calls to the default cost-models with the learned model invocations (Step $10$ in Figure~\ref{fig:resplan}) to predict the exclusive cost of an operator, which is then combined with costs of children operators, similar to how default cost models combine costs. Moreover, since there is an operator model  and a combined model for every physical operator, \system can cover all possible query plans, and can even generate a plan unseen in training data.}
All the features that learned models need are available during query optimization. However, we do not reuse the partition count derived by the default cost model, rather we try to find a more optimal partition count (steps $3$, $4$, $9$) as it drives the query latency. We discuss the problem with existing partition count selection and our solution below.
}

\vspace{-0.4cm}
\subsection{Resource-aware Query Planning}
\label{sec:resourceopt}
\vspace{-0.1cm}
The degree of parallelism (i.e., the number of machines or containers allocated for each operator) is a key factor in determining the runtime of queries in massively parallel databases~\cite{raqo}, which implicitly depends on the partition count. This makes  partition count as an important feature in determining the cost of an operator \papertext{(as noted in Figures~\ref{fig:peroperatorsubgraph})}
\techreport{(as noted in Figures~\ref{fig:peroperatorsubgraph}--\ref{fig:mergedfeatures_imp})}.

Unfortunately, in existing optimizers, the partition count is not explored for all operators, rather partitioning operators (e.g., Exchange for stage $2$ in Figure~\ref{fig:resplan_example}) set the partition count for the entire stage based on their local statistics~ \cite{yint2018bubble}. The above operators on the same stage simply derive the partition count set by the partitioning operator. For example, in stage $2$ of Figure~\ref{fig:resplan_example}, Exchange sets a partition count to $2$ as it results in its smallest local cost (i.e., $15$).
The operators above Exchange (i.e., Reduce and Output) derive the same partition count, resulting in a total cost of $305$ for the entire stage.
However, we can see that the partition count of $16$ results in a much lower overall cost of $125$, though it's not locally optimal for Exchange. Thus, \emph{not optimizing the partition count for the entire stage results in a sub-optimal plan}.

To address this, we explore the partition counts during query planning, by making query planning resource-aware.
Figures~\ref{fig:resplan} and~\ref{fig:resplan_example} illustrate our resource-aware query planning approach.
We introduce the notion of a \emph{resource-context}, within the optimizer-context, for tracking costs of partitions across operators in a stage. Furthermore, we add a \emph{partition exploration} step, where each 
physical operator attaches  a list of learned costs for different partition counts to the resource context (step 3 in Figure~\ref{fig:resplan}). For example, in Figure~\ref{fig:resplan_example}, the  resource-context for stage 2  shows the learned cost for different partition counts for each of the three operators. On reaching the stage boundary, the partitioning operator Exchange performs \emph{partition optimization} (step $9$ in Figure~\ref{fig:resplan}) to set its local partition count to $16$, that results in the lowest total cost of $125$ across for the stage.
Thereafter, the higher level operators simply derive the selected partition count (line 8 in Figure~\ref{fig:resplan}), like in the standard query planning, and estimate their local costs using learned models. Note that when a partition  comes as required property~\cite{cascades95} from upstream operators, we set the partition count to the required value without any exploration (Figure~\ref{fig:resplan} step 2). 

\scope{} currently does not allow varying other resources, therefore we focus only on partition counts in this work. However, the resource-aware query planning with the three new abstractions to Cascades framework, namely the resource-context, partition-exploration, and partition-optimization, is general enough to incorporate additional resources such as memory sizes, number of cores, VM instance types, and other infrastructure level decisions to jointly optimize for both plan and resources. Moreover, our proposed extensions can also be applied to other big data systems such as Spark, Flink, and Calcite that use variants of Cascades optimizers and follow a similar top-down query optimization as \scope{}.

Our experiments in Section~\ref{sec:experiments} show 
that the resource-aware query planning not only generates better plans in terms of latency, but also leads to resource savings. However, the challenge is that estimating the cost for every single partition count for each operator in the query plan can explode the search space and make query planning infeasible.
We discuss our approach to address this next.

\vspace{-0.3cm}
\subsection{Efficient Resource Exploration}

We now discuss two techniques for efficiently exploring the partition counts {in \system}, without exploding the search space.

\stitle{Sampling-based approach.} Instead of considering every single partition count, one option is to consider a uniform sample over the set of all possible containers for the tenant or the cluster. However, the relative change in partition is more interesting when considering its influence on the cost, e.g., a change from $1$ to $2$ partitions influences the cost more than a change from $1200$ to $1210$ partitions.
Thus, we sample partition counts in a geometrically increasing sequence, where a sample  $x_{i+1}$ is derived from previous sample $x_{i}$ using: $x_{i+1} = \ceil*{x_{i} +x_{i}/s}$ with $x_0=1$, $x_1=2$. Here, $s$ is a skipping coefficient that decides the gap between successive samples. A large $s$ leads to a large number of samples and more accurate predictions, but at the cost of higher model look-ups and prediction time.

\techreport{
	\begin{figure}
		\centerline {
			\hbox{\resizebox{\columnwidth}{!}{\includegraphics{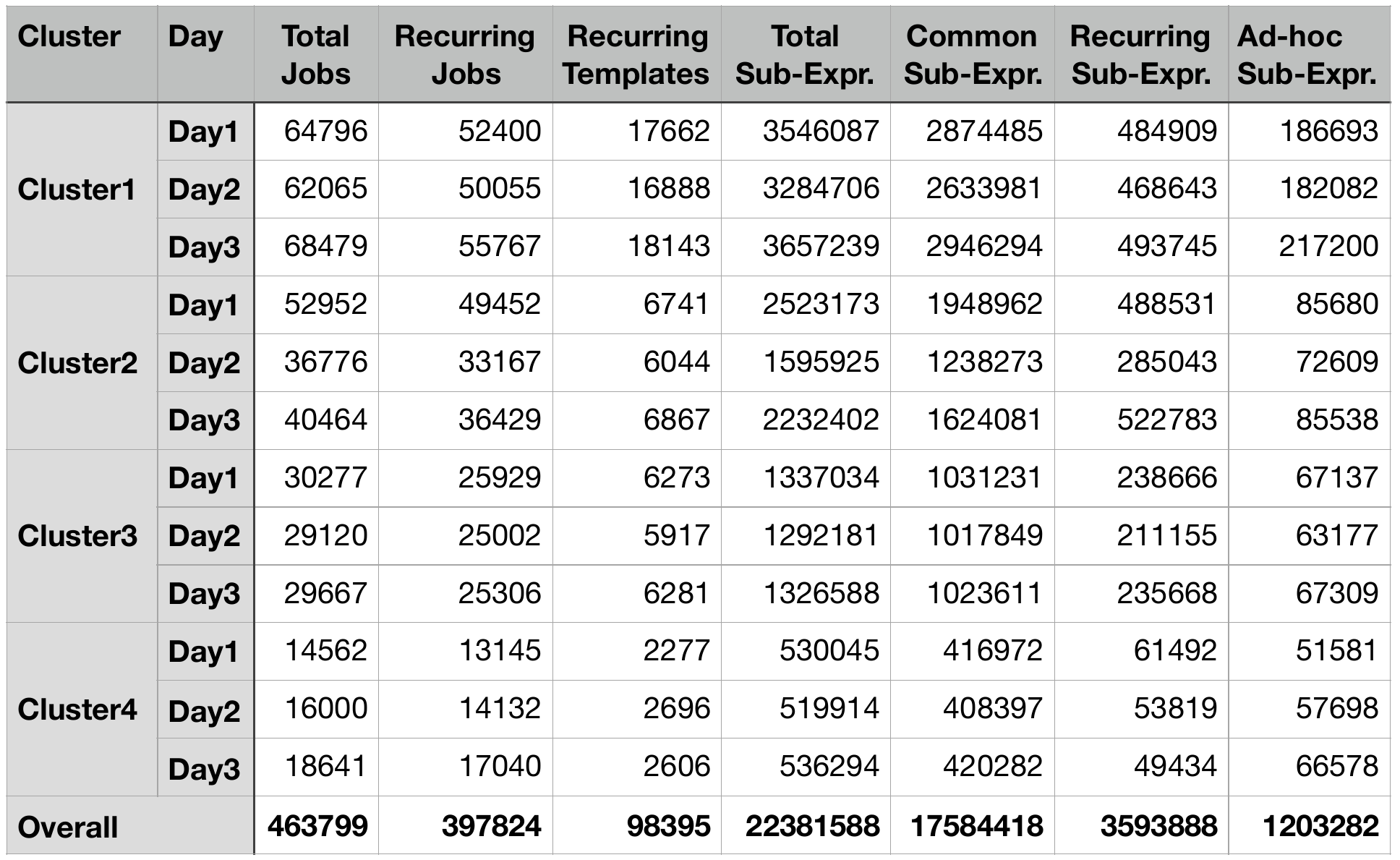}}}}
		\vspace{-10pt}
		\caption{\shep{Workload consisting of $0.5$ million jobs from $4$ different production clusters over $3$ days}}
		\label{fig:workload_summary}
		\vspace{-15pt}
	\end{figure}
}

\stitle{Analytical approach.}
We reuse the individual learned models to directly model the relationship between the partition count and the cost of an operator.
The key insight here is that only features where partition count is present are relevant for partition exploration, while the rest of the features can be considered constants since their values are fixed during partition exploration.
Thus, we can express operator cost as follows: 
$cost \propto \frac{(\theta_1*I+\theta_2*C + \theta_3*I*C)}{P} + \theta_c*P$ where $I$, $C$, and $P$ refer to input cardinality, cardinality and partition count respectively. During optimization, we know $I$, $C$, $I*C$, therefore: $cost \propto \frac{\theta_P}{P} + \theta_c*P$. Extending the above relationship across all operators (say $n$ in number) in a stage, the relationship can be modeled as: $cost \propto \frac{\sum_{i=1}^n \theta_{P_i}}{P} + \sum_{i=1}^n \theta_{C_i}*P$ 

Thus, during partition exploration, each operator calculates $\theta_{P}$ and $\theta_{C}$ and adds them to resource-context, and the partitioning operator selects the optimal partition count by optimizing the above function. There are three possible scenarios: \rev{
(i) ~$\sum_{i=1}^n \theta_{P_i}$ is positive while $\sum_{i=1}^n \theta_{C_i}$ is negative: we can have the maximum number of partitions for the stage since there is no overhead of increasing the number of partitions,} \rev{
(ii)~$\sum_{i=1}^n \theta_{P_i}$ is negative while $\sum_{i=1}^n \theta_{C_i}$ is positive: we set the partition count to minimum as increasing the partition count increases the cost,
}(iii)~$\sum_{i=1}^n \theta_{P_i}$ and $\sum_{i=1}^n \theta_{C_i}$ are either both positive or both negative: we can derive the optimal partition count by differentiating the cost equation with respect to P. 
Overall, for $m$ physical operator, and the maximum possible partition count of  $P_{max}$, the analytical model makes $5\cdot m\cdot log_{\frac{s+1}{s}}P_{max}$ cost model look-ups. Figure~\ref{fig:model_invocations} shows the number of model look-ups for sampling and analytical approaches as we increase the number of physical operators from $1$ to $40$ in a plan. While the analytical model incurs a maximum of $200$ look-ups, the sampling approach can incur several thousands depending on the skipping coefficients.
\techreport{In section~\ref{subsec:partexplore}, we further analyze the accuracy of the sampling strategy with the analytical model on the production workload as we vary the sample size. Our results show that the analytical model is at least $20$ $\times$ more efficient than the sampling approach for achieving the same accuracy. Thus, we use the analytical model as our default partition exploration strategy.}

\section{Experiments}
\label{sec:experiments}

\techreport{
	\begin{figure*}
		\centerline {
			\hbox{\resizebox{\textwidth}{!}{\includegraphics{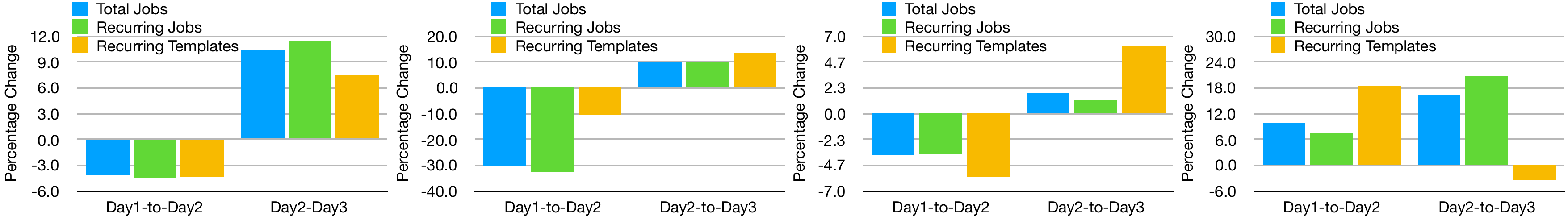}}}}
		\vspace{-5pt}
		\caption{Illustrating workload changes over different clusters and different days. }
		\label{fig:workload_changes}
	\end{figure*}
}

\techreport{
	\begin{figure*}
		\centerline {
			\hbox{\resizebox{\textwidth}{!}{\includegraphics{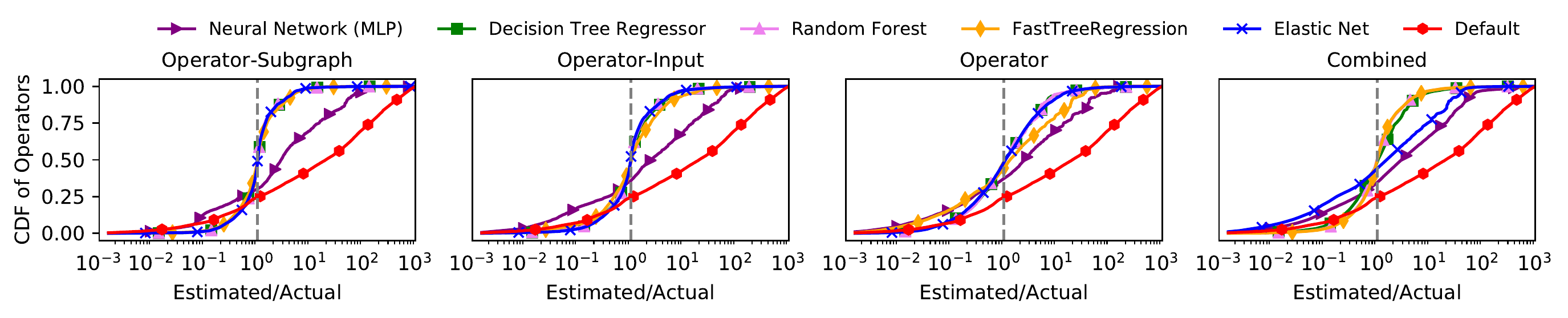}}}}
		\vspace{-0.5cm}
		\caption{\shep{Cross-validation results of ML algorithms for each learned model on Cluster 4 workload}}
		\vspace{-0.3cm}
		\label{fig:ml_cv}
	\end{figure*}
}

\begin{table*}
	\small
	\papertext{\vspace{-5pt}}
	\resizebox{0.88\textwidth}{!}{%
		\begin{tabular}{l|c|c|c|c|c|c|c|c} 
			\hline
			& \multicolumn{4}{c|}{\textbf{All jobs}}  & \multicolumn{4}{c}{\textbf{Ad-hoc jobs}} \\ \hline 
			& Correlation &  Median Error &  95\%tile Error & Coverage &
			Correlation &  Median Error &  95\%tile Error & Coverage \\ \hline 
			\textbf{Default}  & \textbf{0.12} & \textbf{182\%} & 12512\% & \textbf{100\%}
			& \textbf{0.09} & \textbf{204\%} & 17791\% & \textbf{100\%}  \\ \hline
			Op-Subgraph  & 0.86 & 9\% & 56\% & 65\% 
			& 0.81 & 14\% & 57\% & 36\% 		\\ \hline
			Op-Subgraph Approx & 0.85  & 12 \% & 71\%& 82\% 
			&  0.80  & 16 \% & 79\% & 64\% 	 \\ \hline
			Op-Input & 0.81 & 23\% & 90\% & 91\%
			& 0.77 & 26\% & 103\% & 79\% 		 \\ \hline		
			Operator & \textbf{0.76} & 33\% & 138\% & \textbf{100\%} &
			\textbf{0.73} & \textbf{42\%}& 186\% & \textbf{100\%} 
			\\ \hline
			\textbf{Combined} &  \textbf{0.79} & \textbf{21\%} & 112\%  & \textbf{100\%} 
			& \textbf{0.73} & \textbf{29\%} & 134\% & \textbf{100\%}
			\\  \hline
		\end{tabular}%
	}
	\captionof{table}{\small \shep{Breakdown of accuracy and coverage of each learned model for all jobs and ad-hoc jobs separately on Cluster1.}}
	\label{tab:cluster1results}
	\vspace{-20pt}
\end{table*}

In this section, we  present an evaluation of our learned optimizer \system.
For fairness, we feed the same statistics (e.g., cardinality, average row length)  to learned models that are used by the \scope{}  default cost model. Our goals are  five fold:
(i)~\shep{to compare the prediction accuracy of our learned cost models over all jobs as well as over only ad-hoc jobs across multiple clusters},
(ii)~to test the coverage and accuracy of learned cost models over varying test windows,
(iii)~to compare the \system{} cost estimates with those from CardLearner, (iv)\shep{~to explore why perfect cardinality estimates are not sufficient for query optimization},
(iv)~to evaluate the effectiveness of sampling strategies and the analytical approach proposed in Section~\ref{sec:resourceopt} in finding the optimal resource (i.e., partition count),
and
(v)~to analyze the performance of plans produced by \system with those generated from the default optimizer in \system{} using both the production workloads and the TPC-H benchmark
(v)~to understand the training and runtime overheads when using learned cost models.

\techreport{
	\begin{table}
		\small
		\papertext{\vspace{-13pt}}
		\resizebox{\columnwidth}{!}{%
			\begin{tabular}{p{1.12cm}|p{1.3cm}|p{1.3cm}|p{1.3cm}|p{1.3cm}|p{1.3cm}|p{1.3cm}} 
				\hline
				Cluster & \multicolumn{2}{c|}{\textbf{Default (all jobs)}} & \multicolumn{2}{c|}{\textbf{Learned (all jobs)}} &
				\multicolumn{2}{c}{\textbf{Learned (ad-hoc jobs)}} \\ \hline 
				& Correlation & Median Accuracy  
				& Correlation & Median Accuracy 
				& Correlation & Median Accuracy  \\ \hline 
				Cluster 1 & 0.12 & 182\% & 0.79 & 21\% & 0.73 & 29\%\\ \hline
				Cluster 2  & 0.08 & 256\% & 0.77 & 33\% & 0.75 & 40\% \\ \hline
				Cluster 3 & 0.15 & 165\% & 0.83 & 26\% & 0.81 & 38\% \\ \hline
				Cluster 4 & 0.05 & 153\% & 0.74 & 15\% & 0.72 & 26\% \\ \hline
			\end{tabular}%
		}
		\captionof{table}{\small \shep{Pearsion Correlation and Median accuracy of default and combined learned model over all jobs and ad-hoc jobs on each cluster.}}
		\label{tab:medianaccuracy}
		\vspace{-25pt}
	\end{table}
}

\shep{
\vspace{-2pt}
\stitle{Workload.} As summarized in Figure~\ref{fig:workload_summary}, we  consider a large workload trace from $4$ different production clusters comprising of $423$ virtual clusters, with each virtual cluster roughly representing a business unit within Microsoft, and consisting a total of $\approx 0.5$~million jobs over $3$ days, that ran with a total processing time of $\approx 6$~million hours, and use a total of $\approx 1.4$~billion containers. The workload exhibits variations in terms of the load (e.g., more than $3 \times$ jobs on Cluster1 compared to Cluster4), as well as in terms of job properties such as average number of operators per job ($50s$ in Cluster1 compared to $30s$ in Cluster4) and the average total processing time of all containers
per job (around $17$ hours in Cluster1 compared to around $5$ hours in Cluster4). The workload also varies across days from a $30\%$ decrease to a $20\%$ increase of different job characteristics on different clusters\techreport{(Figure~\ref{fig:workload_changes})}. Finally, the workload consists of a mix of both recurring and ad-hoc jobs, with about $7\%-20\%$ ad-hoc jobs on different clusters and different days.
}

\techreport{	
	\begin{figure*}
		\hspace{-0.1cm}
		\centerline {
			\hbox{\resizebox{\textwidth}{!}{\includegraphics{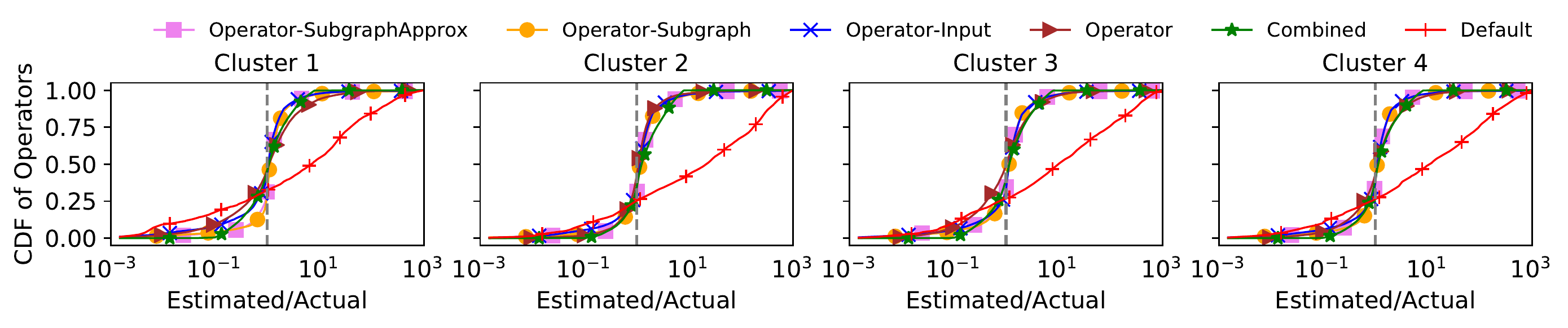}}}}
		\vspace{-25pt}
		\caption{\shep{Accuracy results on all jobs (recurring + ad-hoc) over four different clusters}}
		\vspace{-5pt}
		\label{fig:overallaccuracyresults}	
	\end{figure*} 
	
	\begin{figure*}
		\hspace{-0.1cm}
		\centerline {
			\hbox{\resizebox{\textwidth}{!}{\includegraphics{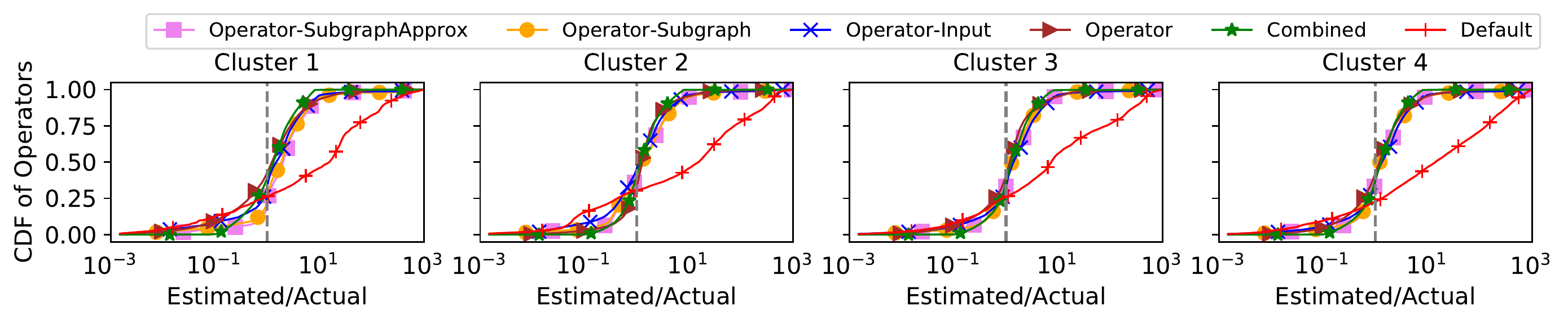}}}}
		\vspace{-25pt}
		\caption{\shep{Accuracy results on only ad-hoc jobs over four different clusters.}}
		\vspace{-5pt}
		\label{fig:nonrecurringaccuracyresults}	
	\end{figure*} 
}

\eat{
	\begin{figure*}
		\vspace{-30pt}
		\centerline {
			\hbox{\resizebox{\textwidth}{!}{\includegraphics{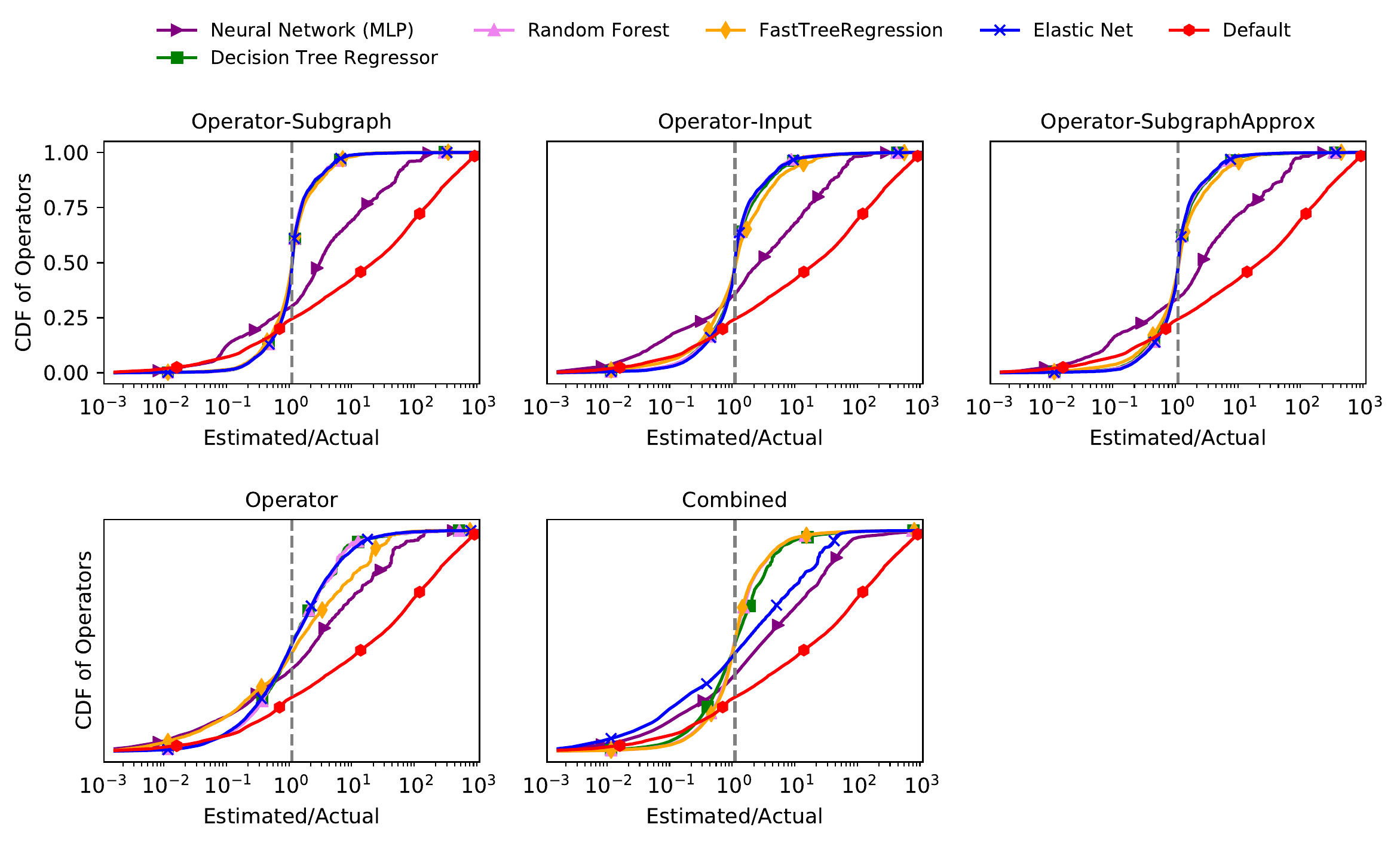}}}}
		\vspace{-0.5cm}
		\caption{{\small Cross-validation results of ML algorithms for each learned model}}
		\label{fig:ml_cv}
	\end{figure*}
}

\techreport{
\vspace{-0.2cm}
\subsection{Cross Validation of ML Models}
\label{sec:evalMLAccuracy}
\shep{
We first compare default cost model with the five  machine learning algorithms discussed in Section~\ref{sec:learningmodel}.
 (i)~Elastic net, a regularized linear-regression model, (ii)~DecisionTree Regressor, (iii)~Random Forest, (iv)~Gradient Boosting Tree, and (v)~ Multilayer Perceptron Regressor (MLP).
Figure~\ref{fig:ml_cv}~(a to d) depicts the $5$-fold cross-validation results for the ML algorithms for  operator-subgraph, operator-input, operator, and combined models respectively. 
We skip the results 
for operator-subgraphApprox as it has similar results to that of operator-input.
We observe that all  algorithms result in better accuracy compared to the default cost model for each of the models. For operator-subgraph and operator-input models, the performance of most of the algorithms (except the neural network) is highly accurate. This is because there are a large number of specialized models, highly optimized for specific sub-graph and input instances respectively. The accuracy degrades as the heterogeneity of model increases from operator-subgraph to operator-input to operator. For individual models, the performances of elastic-net and decision-tree are similar or better than complex models such as neural network and ensemble models. This is because the complex models are more prone to overfitting and noise as discussed in Section~\ref{sec:learningmodel}. For the combined model, the FastTree Regression does better than other models because of its ability to better characterize the space where each model performs well. The operator-subgraph and operator-input models have the highest accuracy (at between $45$\% to $85$\% coverage), followed by the combined model (at $100$\% coverage) and then the operator model (at $100$\% coverage).

}

}

\shep{
\vspace{-10pt}
\subsection{Accuracy}
\label{sec:accuracy}
\papertext{We first compare the accuracy and correlation of learned models to that of the default cost model for each of the clusters.}
\techreport{Next, we first compare the accuracy and correlation of learned models to that of the default cost model for each of the clusters.}
We use the elastic net model for individual learned model and the Fast-Tree regression for the combined model.
\papertext{Cross-validation results for other ML models are presented in Tables 6, 7 and 8, and the CDF distribution of predictions are shown in our technical report~\cite{techreport}.}
We learn the models on day 1 and day 2, and predict on day 3 of the workload as described in Section~\ref{trainfeedbak}.
Table~\ref{tab:medianaccuracy} shows the Pearson correlation and median accuracy of default cost model and the combined model for all jobs and only ad-hoc jobs separately for each of the clusters on day 3. We further show the breakdown of results for each of the individual models on cluster 1 jobs in Table~\ref{tab:cluster1results}. \techreport{ Figure~\ref{fig:overallaccuracyresults} and Figure~\ref{fig:nonrecurringaccuracyresults} show the CDF distribution for estimated vs actual ratio of predicted costs for each of the learned models over different clusters.}

\stitle{All jobs.} We observe that learned models result between $8 \times$ to $10 \times$ better accuracy and between $6 \times$ to $14\times$ better correlation compared to the default cost model across the $4$ clusters. For operator-subgraph, the performance is highly accurate (9\% median error and .86 correlation), but at lower coverage of 65\%. This is because there are a large number of specialized models, highly optimized for specific sub-graph instances. As the coverage increases from operator-subgraphApprox to operator-input to operator, the accuracy decreases. Overall, the combined model is able to provide the best of both worlds, i.e., accuracy close to those of individual models and $100$\%  coverage like that of the operator model. The $50^{th}$ and the $95^{th}$ percentile errors of the combined model are about $10$$\times$ and $1000$$\times$ better than the default \scope{} cost model. These results show that it is possible to learn highly accurate cost models from the query workload. 

\begin{figure*}
	\vspace{-0.05cm}
	\centerline {
		\hbox{\resizebox{\textwidth}{!}{\includegraphics{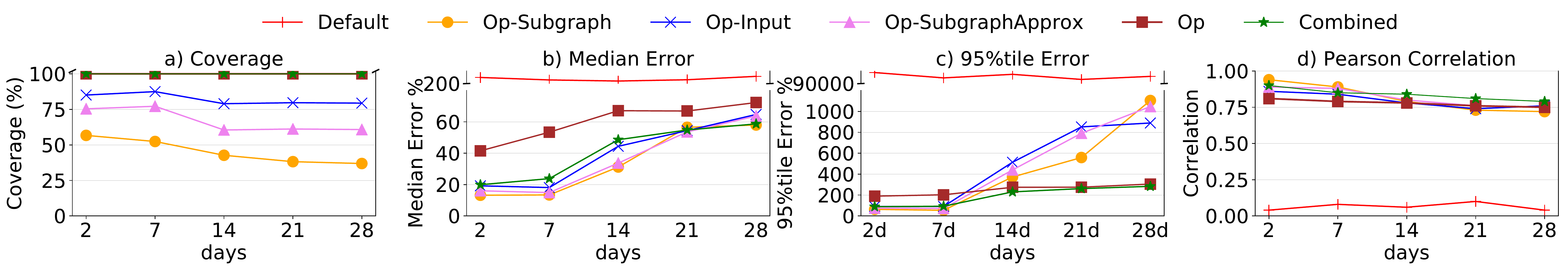}}}}
	\vspace{-0.5cm}
	\caption{{\small Coverage and accuracy with respect to default optimizer over 1 month} }
	\vspace{-0.4cm}
	\label{fig:perfovermonth}	
\end{figure*}

\stitle{Only ad-hoc Jobs.} Interestingly, \emph{the accuracy over ad-hoc jobs drop slightly but it is still close to those over all jobs (Table~\ref{tab:medianaccuracy} and Table~\ref{tab:cluster1results})}. This is because: (i) ad-hoc jobs can still have one or more similar subexpressions as other jobs (e.g., they might be scanning and filtering the same input before doing completely new aggregates), which helps them to leverage the subgraph models learned from other jobs. This can be seen in  Table~\ref{tab:cluster1results} --- the sub-graph learned models still have substantial coverage of subexpression on ad-hoc jobs. For example, the coverage of $64\%$ for Op-Subgraph Approx model means that $64\%$ of the sub-expressions on ad-hoc jobs had matching Op-Subgraph Approx learned model. (ii)  Both the operator and the combined models are learned on a per-operator basis, and have much lower error ($42\%$ and $29\%$) than the default model ($182\%$). This is because the query processor, the operator implementations, etc. still remain the same, and their behavior gets captured in the learned cost models. Thus, even if there is no matching sub-expression in ad-hoc jobs, the operator and the combined models still result in better prediction accuracy.

}

\vspace{-0.3cm}
\subsection{Robustness}
\label{sec:exp-robustness}
We now look at the robustness (as defined in Section~\ref{sec:intro}) of learned models in terms of accuracy, coverage, and retention over a month long test window on cluster 1 workloads.

\stitle{Coverage over varying test window.} Figure~\ref{fig:perfovermonth}a depicts the coverage of different subgraph models as we vary the test window over a duration of 1 month. 
The coverage of per-operator and combined model is always $100\%$ since there is one model for every physical operator. The coverage of per-subgraph models, strictest among all, is about $58\%$ after $2$ days, and decreases to $37\%$ after $28$ days. Similarly, the coverage of per-subgraphApprox ranges between $75\%$ and $60\%$. The per-operator-input model, on the other hand remains stable between $78\%$ and $84\%$.

\stitle{Error and correlation over varying test window.} Figures~\ref{fig:perfovermonth}b and~\ref{fig:perfovermonth}c depict the median and $95\%$ile error percentages respectively over a duration of one month. While the median error percentage of learned models improves on default cost model predictions by $3$-$15$x, the $95\%$ile error percentage is better by over three orders of magnitude. For specialized learned models, the error increases slowly over the first two weeks and then grows much faster due to decrease in the coverage. 
Moreover, the $95\%$ile error of the subgraph models grows worse than their median error. 
Similarly, in Figure~\ref{fig:perfovermonth}d, we see that predicted costs from learned models are much more correlated with the actual runtimes, having high Pearson correlation (generally between $0.70$ and $0.96$), compared to default cost models that have a very small correlation (around $0.1$). A high correlation over a duration month of $1$ month shows that learned models can better discriminate between two candidate physical operators. Overall, based on these results, we believe re-training every $10$ days should be acceptable, with a median error of about $20\%$, $95\%$ error of about $200\%$, and Pearson correlation of around $0.80$. 

\stitle{Robustness of the combined model}. From Figure~\ref{fig:perfovermonth}, we note that the combined model (i) has 100\% coverage, (ii) matches the accuracy of best performing individual model at any time (visually illustrated in Figure~\ref{fig:cov_accuracy_heatmap}), while having a high correlation ($>$ 0.80) with actual runtimes, and (iii) gives relatively stable performance with graceful degradation in accuracy over longer run. Together, these results show that the combined model in \system{} is indeed robust.

\begin{figure*}
	\vspace{-25pt}
	\begin{minipage}{.34\textwidth}
		\centerline {
			\hbox{\resizebox{0.8\columnwidth}{!}{\includegraphics{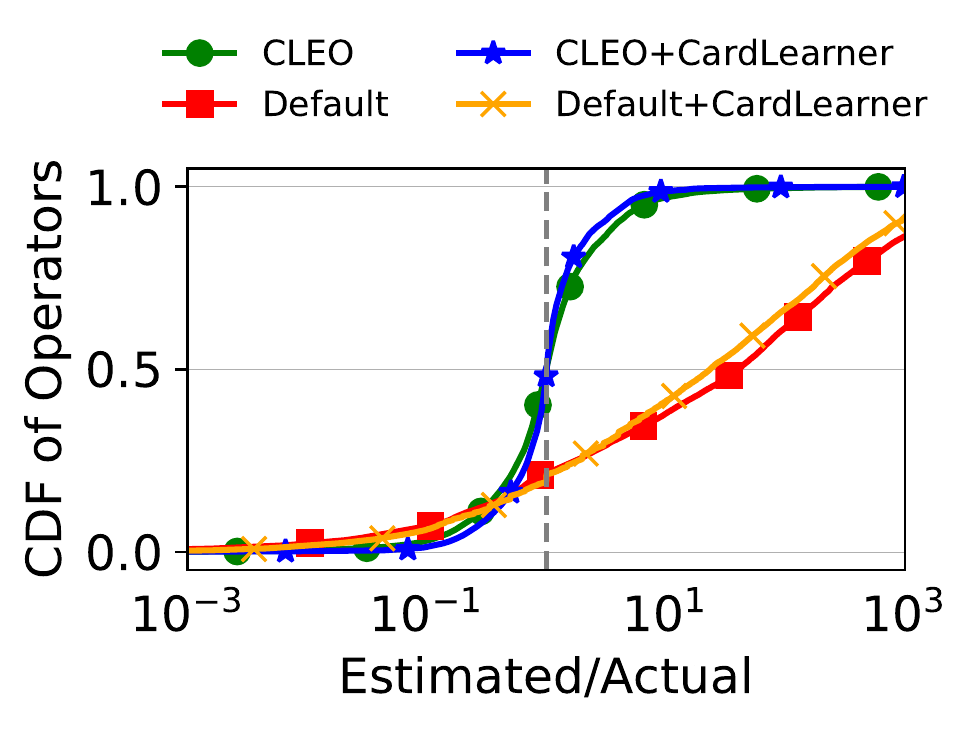}}}}
		\vspace{-15pt}
		\caption{\rev{\small Comparison with CardLearner}}
		\label{fig:cardlearner}
	\end{minipage}
	\hspace{0.1cm}
	\begin{minipage}{.33\textwidth}
		\vspace{-10pt}
		\centerline {
			\hbox{\resizebox{1.05\columnwidth}{!}{\includegraphics{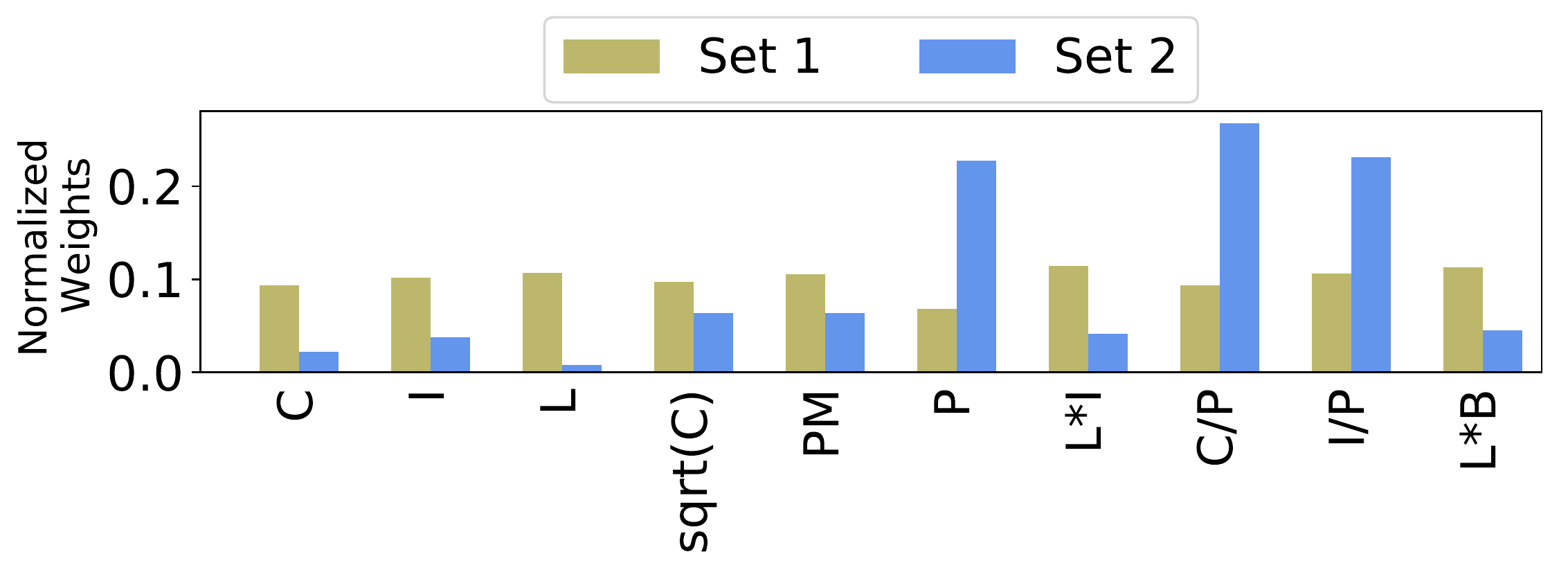}}}}
		\vspace{-12pt}
		\caption{\shep{Hash join operator having different weights over different sets of sub-expressions.}}
		\vspace{-12pt}
		\label{fig:differrentweights}
	\end{minipage}
	\hspace{0.2cm}
	\begin{minipage}{.26\textwidth}
		\vspace{-15pt}
		\centerline {
			\hbox{\resizebox{0.9\columnwidth}{!}{\includegraphics{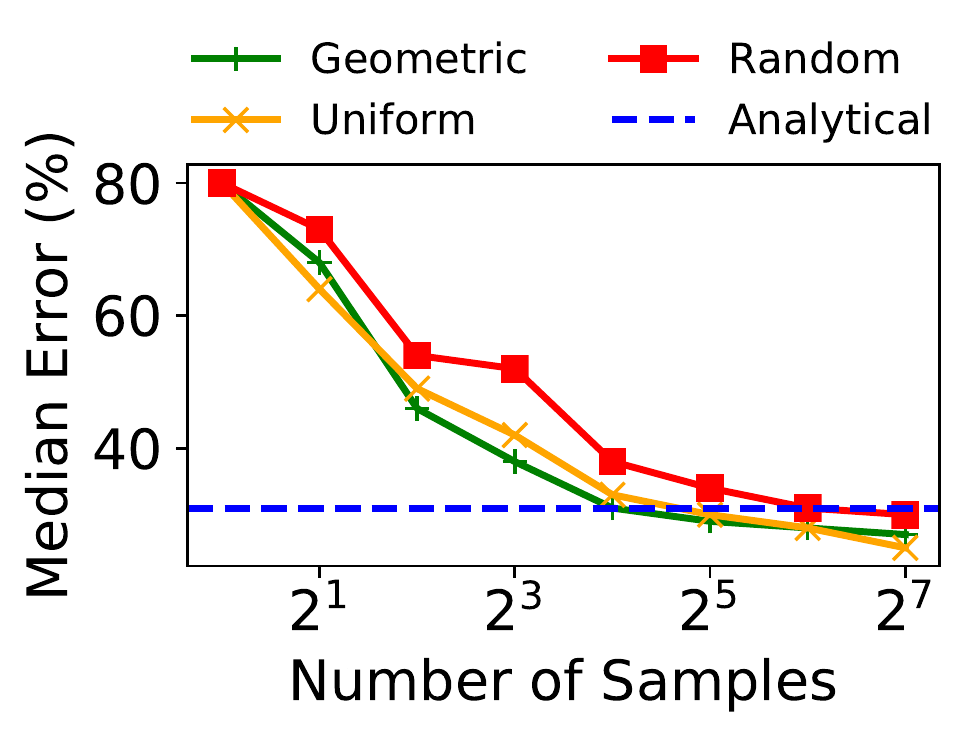}}}}
		\vspace{-12pt}
		\caption{\rev{\small Partition Exploration Accuracy vs. Efficiency }}
		\label{fig:resourceopt}
		\vspace{-15pt}
	\end{minipage}
	\vspace{-2pt}
\end{figure*}

\vspace{-5pt}
\subsection{Impact of Cardinality}
\label{sec:cardLearner} 
	\vspace{-5pt}
\stitle{Comparison with Cardlearner.} Next, we compare the accuracy and the Pearson correlation of \system and the default cost model with CardLearner~\cite{cardLearner}, a learning-based cardinality estimation that employs a Poisson regression model to improve the cardinality estimates, but uses the default cost model to predict the cost. For comparison, we considered jobs from a single virtual cluster from cluster 4, consisting of about $900$ jobs. 
While \system and the default cost model use the cardinality estimates from the \scope{} optimizer, we additionally consider a variant (\system+CardLearner) where \system uses the cardinality estimates from CardLearner. Overall, we observe that the median error of default cost with CardLearner (211\%) is better than that of default cost model alone (236\%) but much still worse than that of \system's (18\%) and \system + CardLearner (13\%). Figure~\ref{fig:cardlearner} depicts the CDF of estimated and actual costs ratio, where we can see that  by learning costs, \system significantly reduces both the under-estimation as well as over-estimation, while CardLearner only marginally improves the accuracy for both default cost model and \system. Similarly, the Pearson correlation of CardLearner's  estimates (.01) is much worse than that of \system's ($0.84$) and \system + CardLearner ($0.86$). These results are consistent with our findings from Section~\ref{sec:motivation} where we show that fixing the cardinality estimates is not sufficient for filling the wide gap between the estimated and the actual costs in \scope{}-like systems. 
Interestingly, we also observe that the Pearson correlation of CardLearner is marginally less than that of the default cost model (0.04) inspite of better accuracy, which can happen when a model makes both over and under estimation over different instances of the same operator.

\shep{
\vspace{-2pt}
		
\stitle{Why cardinality alone is not sufficient?} To understand why fixing cardinality alone is not enough, we performed the following two micro-experiments on a subset of jobs, consisting of about $200$ K subexpressions from cluster 4.

\stitle{1. The need for a large number of features.} First, starting with perfect cardinality estimates (both input and output) as only features, we fitted an elastic net model to find / tune the best weights that lead to minimum cost estimation error (using the loss function described in Section~\ref{sec:learningmodel}). Then, we incrementally added more and more features and also combined feature with previously selected features, retraining the model with every addition. Figure~\ref{fig:cumfeatureError} shows the decrease in cost model error as we cumulatively add features from left to right. We use the same notations for features as defined in Table~\ref{tab:basicfeatures} and Table~\ref{tab:derivedfeatures}.

\begin{figure}[h]
	\centerline {
		\hbox{\resizebox{\columnwidth}{!}{\includegraphics{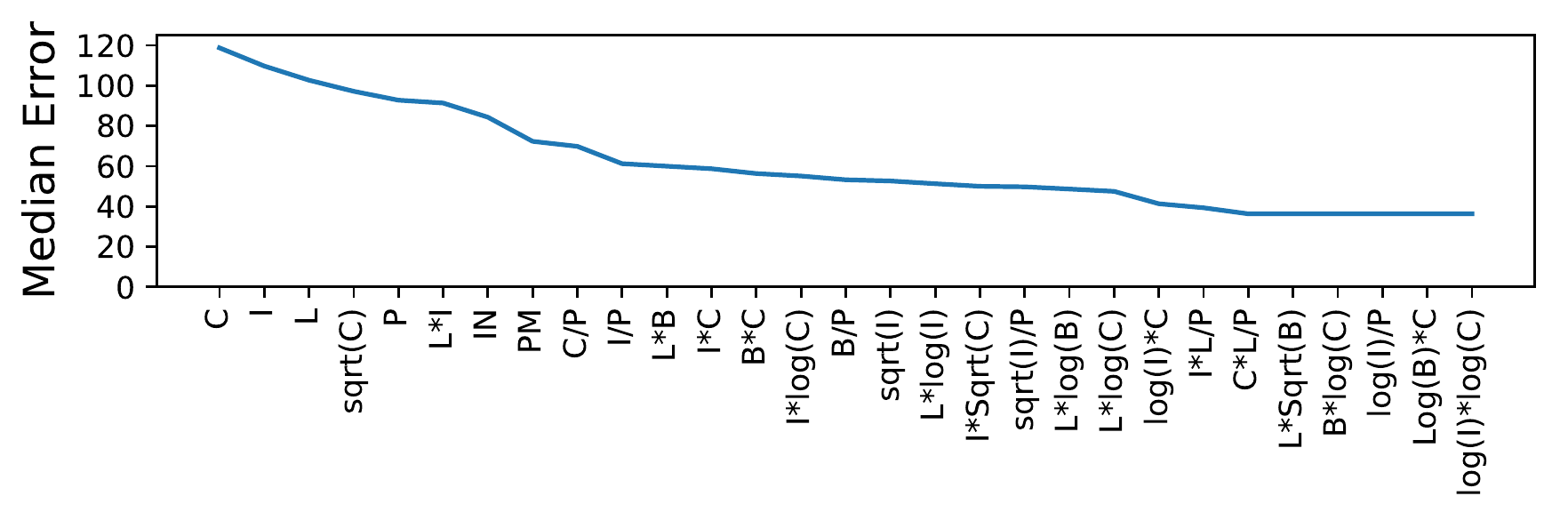}}}}
	\vspace{-14pt}
	\caption{\small \shep{Impact on median error as we cumulatively add features from left to right. The first two features are perfect output (C) and input (I) cardinalities.}}
	\vspace{-8pt}
	\label{fig:cumfeatureError}
\end{figure}

 We make the following observations.
\vspace{-5pt}
\begin{packed_enum}
	\item When using only perfect cardinalities, the median error is about $110\%$. However, when adding more features, we observe that the error drops by more than half to about $40\%$.
	This is because of the more complex environments and queries in big data processing that are very hard to cost with just the cardinalities (as also discussed in Section 2.3). 
	
	\item Deriving additional statistics by transforming (e.g., sqrt, log, squares) input and output cardinalities along with other features helps in reducing the error. However, it is hard to arrive at these transformations in hand coded cost models.
	
	\item We also see that features such as parameter (PM), input (IN), and partitions (P) are quite important, leading to sharp drops in error. While partition count is good indicator of degree of parallelism (DOP) and hence the runtime of job; unfortunately query optimizers typically use a fixed partition count for estimating cost as discussed in Section 5.2. 
	
	\item Finally, there is a pervasive use of unstructured data (with schema imposed at runtime) as well as custom user code (e.g., UDFs) that embed arbitrary application logic in big data applications.
	It is hard to come up with generic heuristics, using just the cardinality, that effectively model the runtime behavior of all possible inputs and UDFs.
\end{packed_enum}

\vspace{-5pt}
\stitle{2. Varying optimal weights.} We now compare the optimal weights of features of physical operators when they occur in different kinds of sub-expressions. 
We consider the popular hash join operator and identified two sets of sub-expressions in the same sample dataset:
(i)~hash join appears on top of two scan operators, and 
(ii)~hash join appears on top of two other join operators, which in turn read from two scans.

Figure~\ref{fig:differrentweights} shows the weights of the top $10$ features of the hash join cost model for the two sets.
We see that the  partition count is more influential for set 2 compared to set 1. This is because there is more network transfer of data in set 2 than in set 1 because of two extra  joins. Setting the partition count that leads to minimum network transfer is therefore important. On the other hand, for jobs in set 1, we notice that hash join typically sets the partition count to the same value as that of inputs, since that minimizes repartitioning. Thus, partition count is less important in set 1. 
To summarize, even when cardinality is present as a raw or derived feature, its relative importance is instance specific (heavily influenced by the partitioning and the number of machines) and hard to capture in a static cost model.
}

\eat{
\begin{figure}
	\centerline {
		\hbox{\resizebox{0.65\columnwidth}{!}{\includegraphics{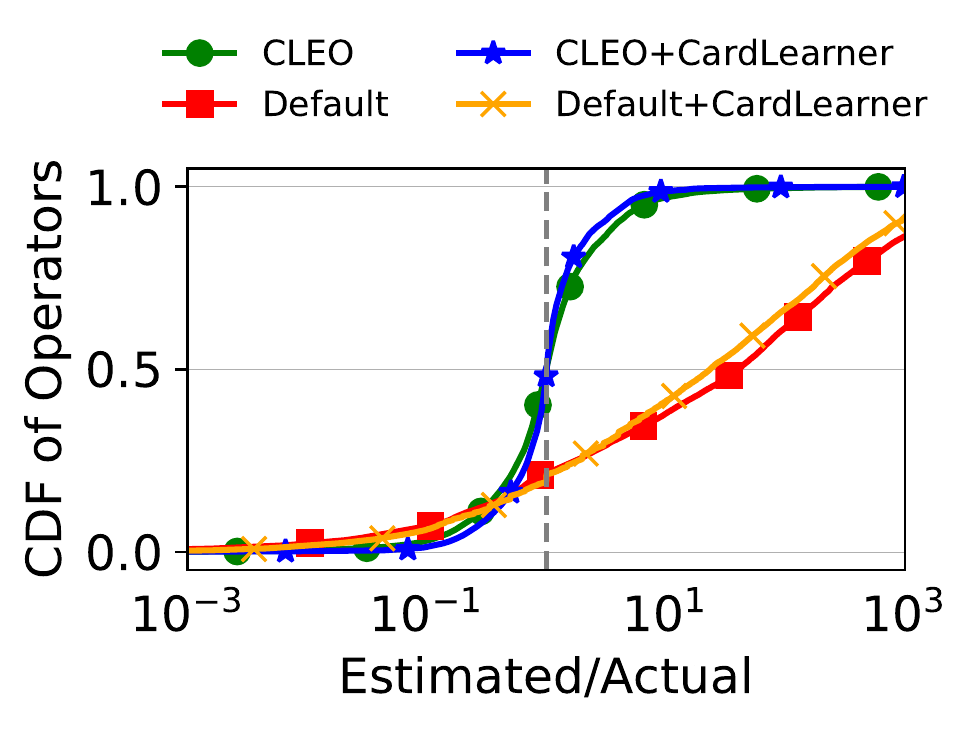}}}}
	\vspace{-10.pt}
	\caption{\rev{Comparison with CardLearner}}
	
	\label{fig:cardlearner}
	\vspace{-10.pt}
\end{figure}
}

\begin{figure*}
	\vspace{1pt}
	\hspace{-0.1cm}
	\begin{subfigure}{0.33\textwidth}
		\centerline {
			\hbox{\resizebox{\columnwidth}{!}{\includegraphics{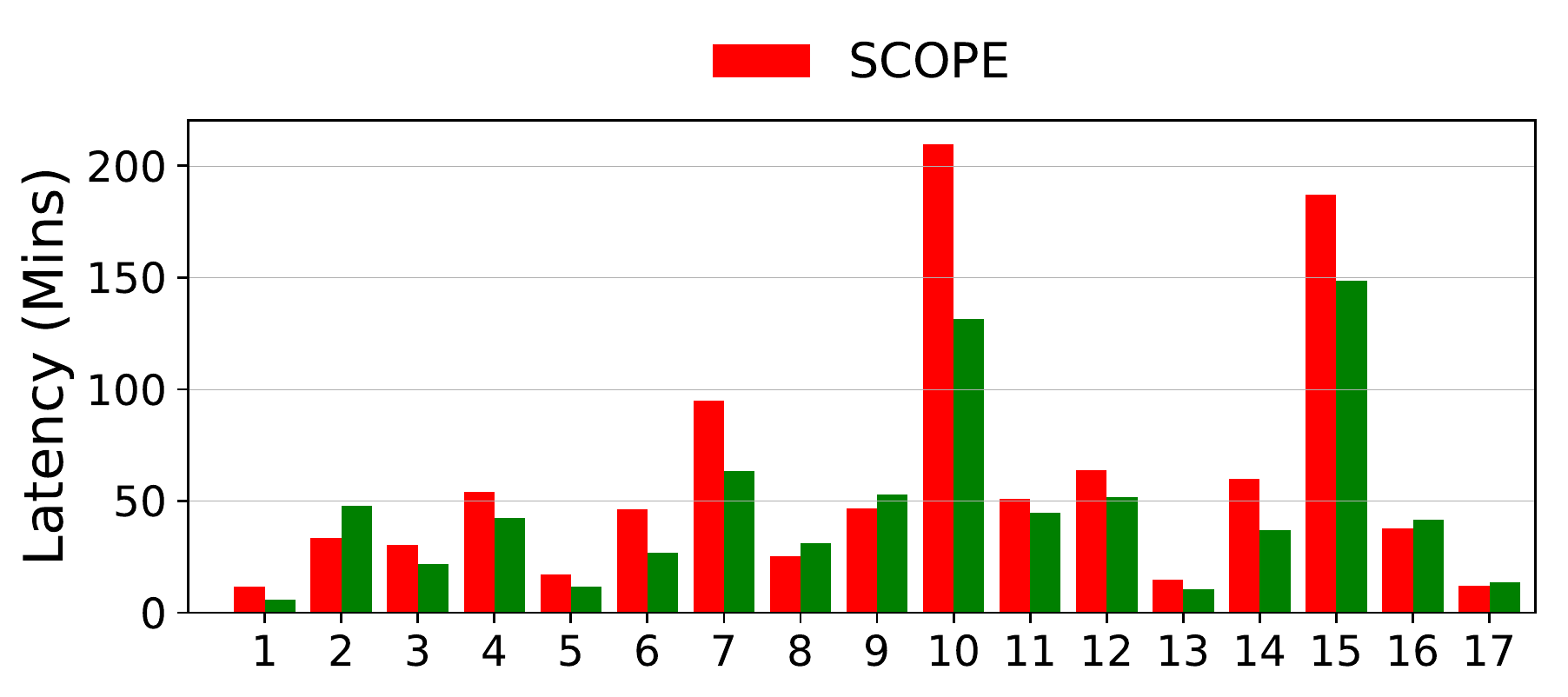}}}}
		\caption{Changes in Latency}
		\label{fig:prodlatency}
	\end{subfigure}
	\hspace{-0.2cm}
	\begin{subfigure}{0.33\textwidth}
		\centerline {
			\hbox{\resizebox{\columnwidth}{!}{\includegraphics{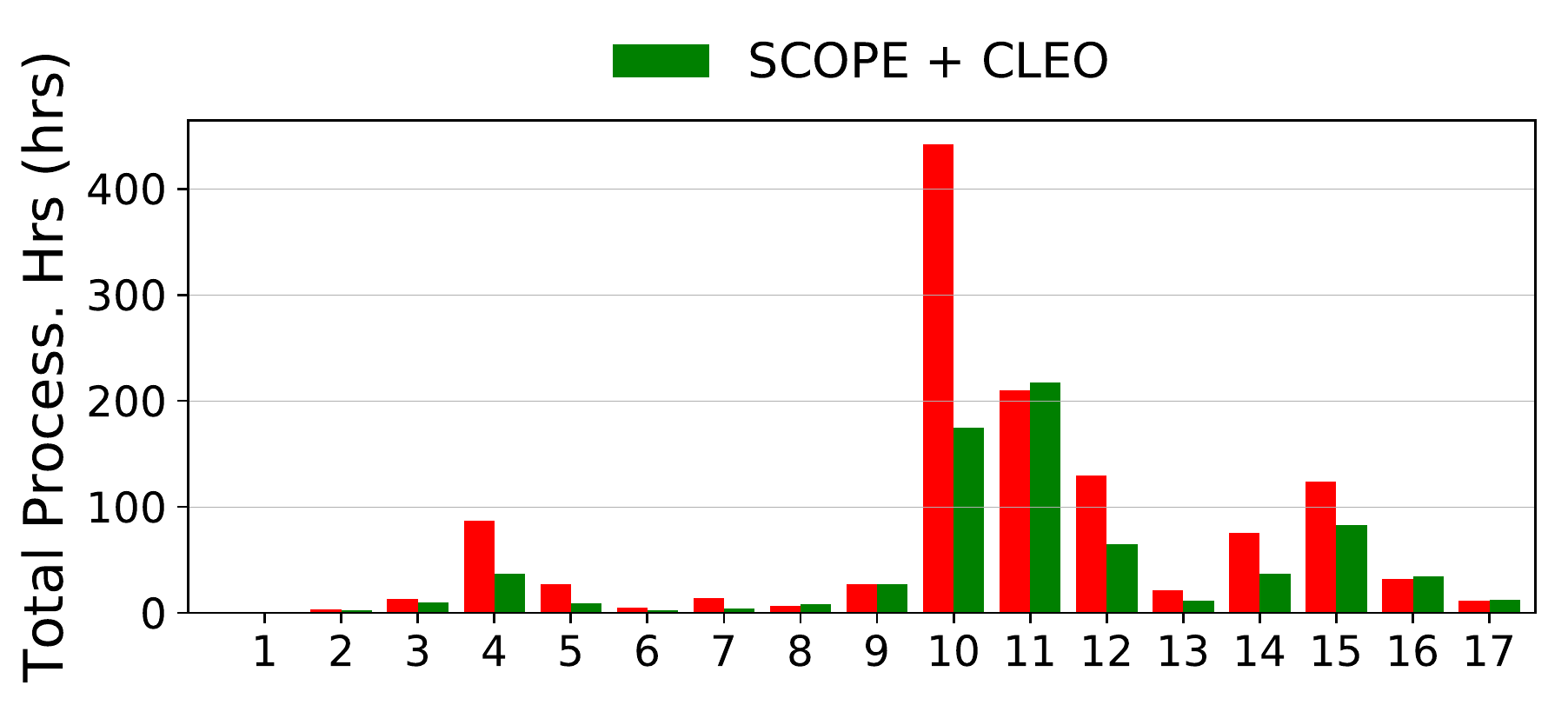}}}}
		\caption{Changes in Total Processing Time}
		\label{fig:prodprocessing}	
	\end{subfigure}
	\vspace{-10pt}	
	\hspace{-0.1cm}
	\begin{subfigure}{0.31\textwidth}
		\centerline {
			\hbox{\resizebox{\columnwidth}{!}{\includegraphics{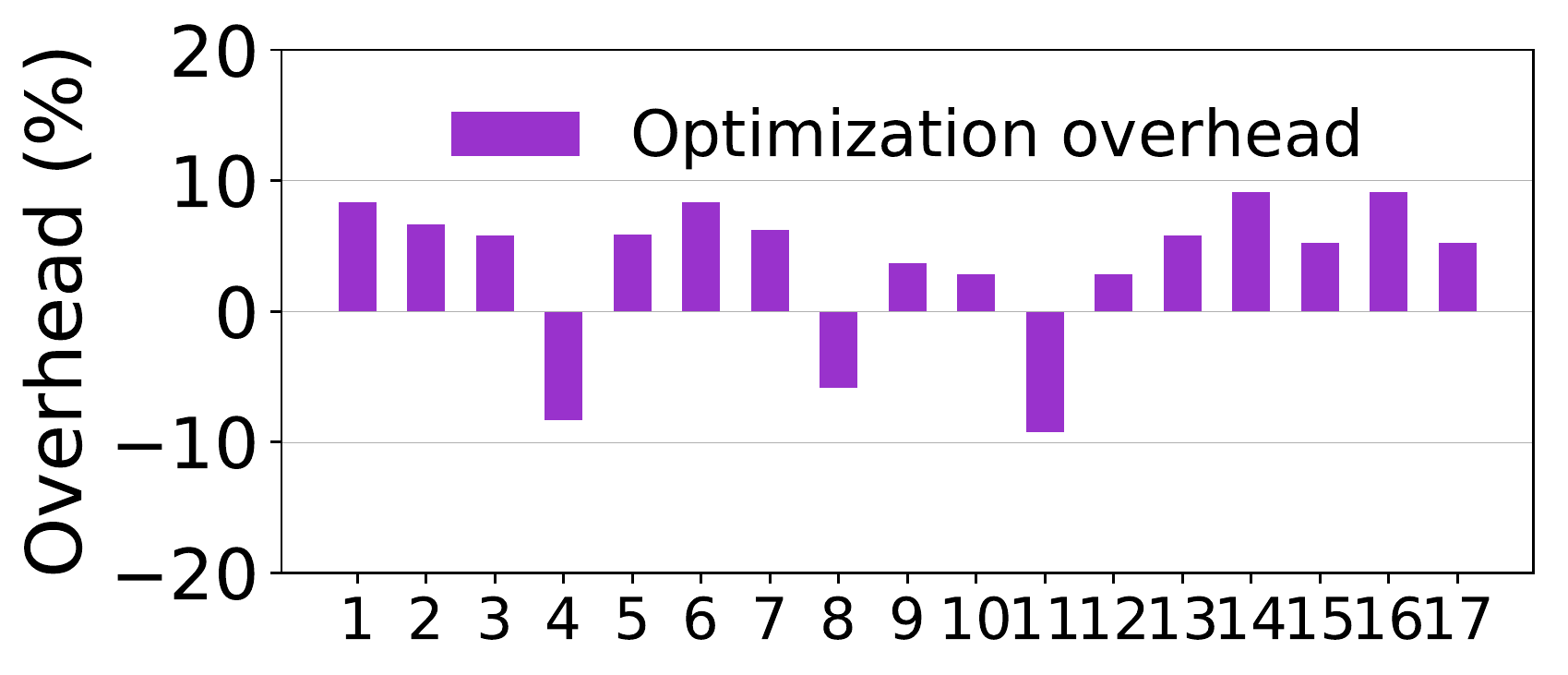}}}}
		\caption{\small \rev{ Optimization Time Overhead}}
		\label{fig:overheadtime}
	\end{subfigure}
	\vspace{-3pt}
	\caption{Performance comparison  on production jobs with changed plans}
	\vspace{-5pt}
	\label{fig:productresults}	
	\vspace{-8pt}
\end{figure*} 

\vspace{-5pt}
\subsection{ Efficacy of Partition Exploration}
\label{subsec:partexplore}

In this section, we explore the effectiveness of partition exploration strategies proposed in Section~\ref{sec:resourceopt} in selecting the partition count that leads to lowest cost for learned models. We used a subset of $200$ sub-expression instances from the production workload from cluster 1, and exhaustively probe the learned models for all partition counts from $0$ to $3000$ (the maximum  capacity of machines on a virtual cluster) to find the most optimal cost.
Figure~\ref{fig:resourceopt} depicts the median error in cost accuracy with respect to the optimal cost for 
(i) three sampling strategies: random, uniform, and geometric as we vary the number of samples of partition counts 
(ii) the analytical model  (dotted blue line) that selects a single partition count.

We note that the analytical model, although approximate, gives more accurate results compared to sampling-based approaches until the sample size of about $15$ to $20$, thereby requiring much fewer model invocations. Further, for a sample size between $4$ to $20$, we see that the geometrically increasing sampling strategy leads to more accurate results compared to uniform and random approaches. This is because it picks more samples when the values are smaller, where costs tend to change more strongly compared to higher values. During query optimization, each sample leads to five learned cost model predictions, four for individual models and one for the combined model. Thus, for a typical plan in big data system that consists of $10$ operators, the sampling approach requires $20*5*10=1000$ model invocations, while the analytical approach requires only $5*10=50$ invocations for achieving the same accuracy. This shows that the analytical approach is practically more effective when we consider both efficiency and accuracy together. Hence, \system uses the analytical approach as the default partition exploration strategy.

\vspace{-0.2cm}
\subsection{Performance}
\label{sec:perf}
We split the performance evaluation of \system into three parts: (i)~the runtime performance over production workloads, (ii)~a case-study on the TPC-H benchmark~\cite{poess2000new}, explaining in detail the plan changes caused by the learned cost models, and (iii)~the overheads incurred due to the learned models. For these evaluations, we deployed a new version of the \scope{} runtime with \system optimizer (i.e., \scope{}+\system ) on the production clusters and compared the performance of this runtime with the default production \scope{} runtime. \rev{We used the analytical approach for partition exploration.}

\vspace{-0.2cm}
\subsubsection{Production workload} 
Since production resources are expensive, we selected a subset of the jobs, similar to prior work on big data systems~\cite{cardLearner}, as follows.
We first recompiled all the jobs from a single virtual cluster from cluster 4 with \system and found that $182$ (22\%) out of $845$ jobs had plan changes when partition exploration was turned off, while $322$ ($39$\%) had plan changes when we also applied partition exploration. For execution, we selected jobs that had at least one change in physical operator implementations, e.g, for aggregation (hash vs stream grouping), join (hash vs merge), addition or removal of grouping, sort or exchange operators. We picked $17$ such jobs and executed them with and without \system over the same production data, while redirecting the output to a dummy location, similar to prior work~\cite{rope}.

Figure~\ref{fig:prodlatency} shows the end-to-end latency for each of the jobs, compared to their latency when using the default cost model.
We see that the learned cost models improve latency in 
70\% (12 jobs) cases, while they degrade latency in the remaining 30\% cases.
Overall, the average improvement across all jobs is $15.35$\%, while the cumulative latency of all jobs improves by $21.3$\%.
Interestingly,  \system was able to improve the end-to-end latency for $10$ out of $12$ jobs with less degree of parallelism (i.e., the number of partitions).
This is in contrary to the typical strategy of scaling out the processing by increasing the degree of parallelism, which does not always help. Instead, resource-awareness plan selection can reveal more optimal combinations of plan and resources.
To understand the impact on resource consumption, Figure~\ref{fig:prodprocessing} shows the total processing time (CPU hours). 
Learned cost models reduce the 
Overall, we see the total processing time reducing by $32.2$\%  on average and $40.4\%$ cumulatively across all 17 jobs--- {\it a significant operational cost saving in \rev{big data} computing environments}.

Thus, the learned cost models could reduce the overall costs while still improving latencies in most cases. Below, we dig deeper into the plans changes using the TPC-H workload.

\eat{
The typical strategy to improve latency in massively parallel systems is to scale out the processing, i.e., increase the degree of parallelism by increasing the number of partitions.
Higher parallelism increases the total processing time, but it is expected to decrease the end to end latency. 
Interestingly, our experiments show that this strategy does not always help. 

As we see from Figure~\ref{fig:prodlatency} and ~\ref{fig:prodprocessing}, \textbf{the resource-aware learned model was able to improve the end-to-end latency for $10$ out of $12$ jobs with less degree of parallelism, resulting in a large amount of reduction (upto $55$\%) in total processing time.
Thus, resource-awareness has the potential to achieve both performance improvement as well as cost savings.}
}

\vspace{-10pt}
\subsubsection{TPC-H workload}
 We generated TPC-H dataset with a scale factor of $1000$, i.e., total input of $1TB$. We ran all $22$ queries $10$ times, each time with randomly chosen different parameters, to generate the training dataset. We then trained our cost models on this workload and feedback the learned models to re-run all $22$ queries. We compare the performance with- and without the feedback as depicted in Figure~\ref{fig:tpchresults} . For each observation, we take the average of $3$ runs.   Overall, $6$ TPC-H queries had plan changes when using resource-aware cost model. Out of these, $4$ changes (Q8, Q9, Q16, Q20) improve latency as well as total processing time, 1 change improves only the latency (Q11), and 1 change (Q17) leads to performance regression in both. We next discuss the key changes observed in the plans.

\noindent
\emph{1. More optimal partitioning.} In Q8, for \texttt{Part} (100 partitions) $\Join_{partKey} $  \texttt{Lineitem} (200 partitions) and in Q9  for \texttt{Part} (100 partitions) $\Join_{partKey}$  (\texttt{Lineitem} 
$\Join$ \texttt{Supplier}) (250 partitions), the default optimizer performs the merge join over 250 partitions, thereby re-partitioning both \texttt{Part} and the other join input into $250$ partitions over the \texttt{Partkey} column. Learned cost models, on other hand, perform the merge join over $100$ partitions, which requires re-partitioning only the other join inputs and not the \texttt{Part} table. Furthermore, re-partitioning $200$ or $250$ partitions into $100$ partitions is cheaper, since it involves partial merge~\cite{bruno2014advanced}, compared to re-partitioning them into $250$ partitions. In $Q16$, for the final aggregation and top-k selection, both the learned and the default optimizer repartition the $250$ partitions from the previous operator's output on the final aggregation key. While the default optimizer re-partitions into $128$ partitions, the learned cost models pick $100$ partitions. The aggregation cost has almost negligible change  due to change in partition counts. However, repartitioning from $250$ to $100$ turns out to be substantially faster than re-partitioning from $250$ to $128$ partitions.

\noindent
\emph{2. Skipping exchange (shuffe) operators.} In Q8, for  \texttt{Part} (100 partitions) $\Join_{partKey} $  \texttt{Lineitem} (200 partitions), the learned cost model performs join over $100$ partitions and thus it skips the costly Exchange operator over the Part table. On the other hand,  the default optimizer creates two Exchange operator for partitioning each input into $250$ partitions.

\noindent
\emph{3. More optimal physical operator:} For both Q8 and Q20, the learned cost model performs join between \texttt{Nations} and \texttt{Supplier} using merge join instead of hash join (chosen by default optimizer). This results in an improvement of $10\%$ to $15\%$ in the end-to-end latency, and $5\%$ to $8\%$ in the total processing time (cpu hours).

\eat{
\begin{figure}
	\hspace{-0.1cm}
	\begin{subfigure}{0.48\columnwidth}
		\centerline {
			\hbox{\resizebox{\columnwidth}{!}{\includegraphics{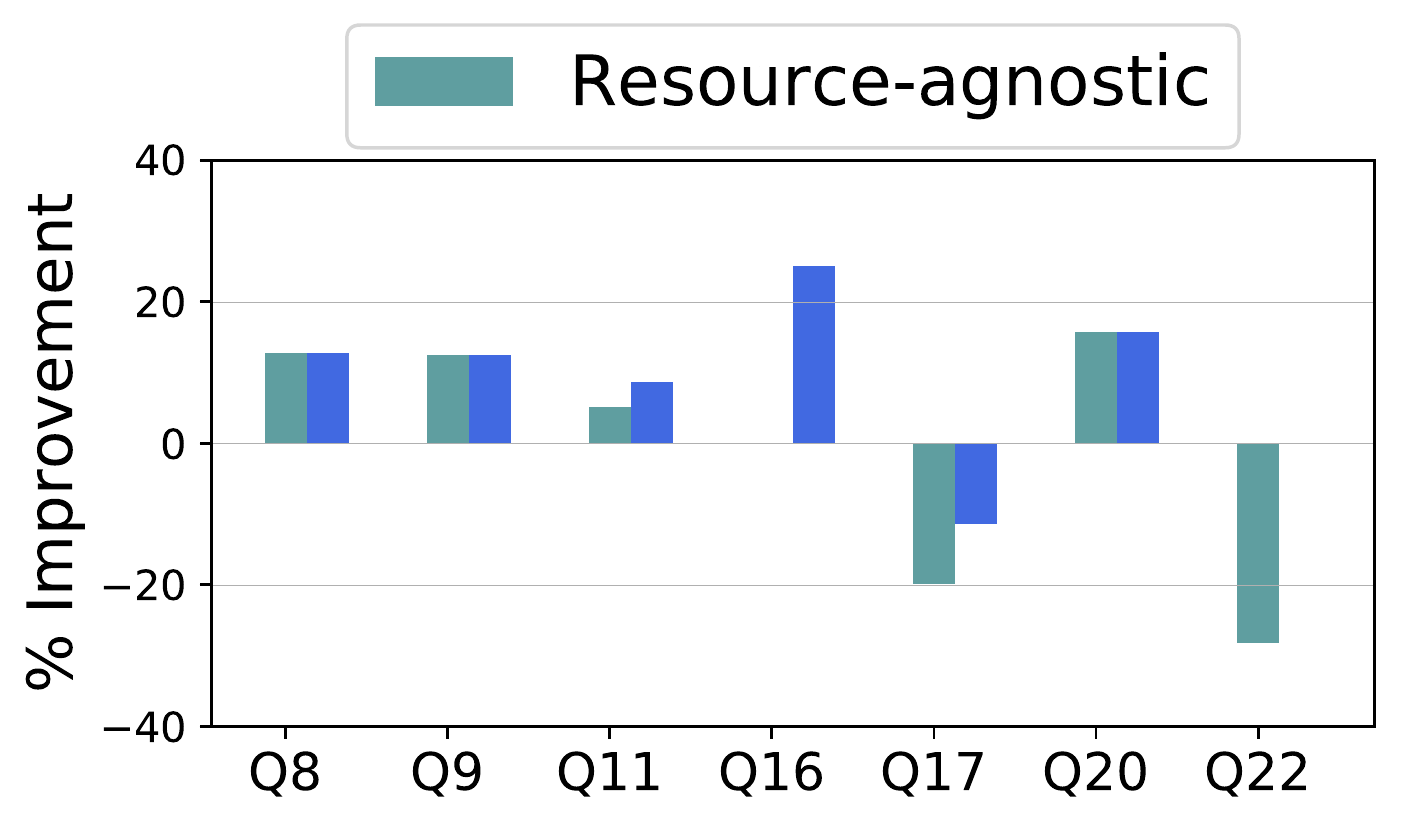}}}}
		\caption{Latency}
		\label{fig:prodlatency}
	\end{subfigure}
	\hspace{-0.2cm}
	\begin{subfigure}{0.48\columnwidth}
		\centerline {
			\hbox{\resizebox{\columnwidth}{!}{\includegraphics{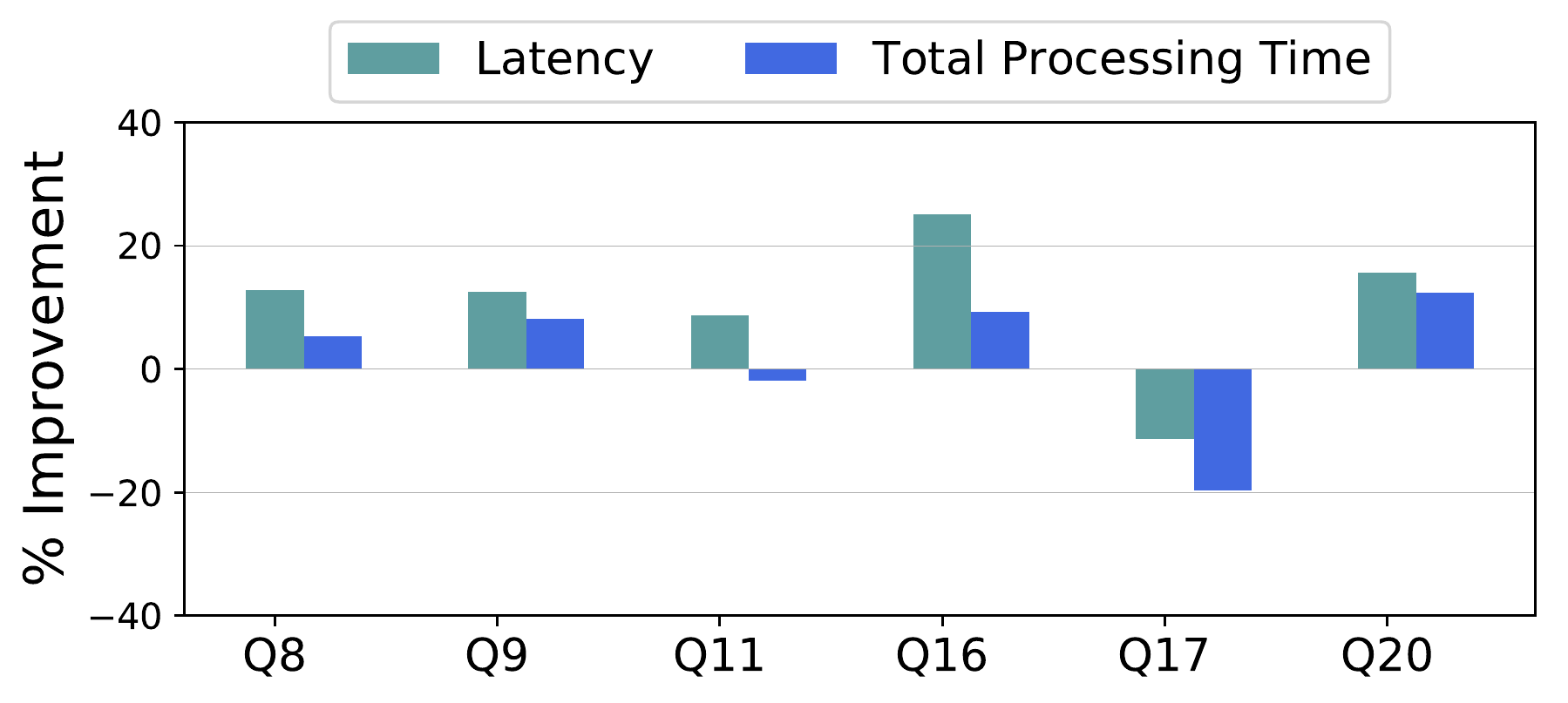}}}}
		\caption{Total Processing Time}
		\label{fig:prodprocessing}	
	\end{subfigure}
	\vspace{-12pt}
	\caption{Performance change with respect to default optimizer for TPC-H queries}
	\vspace{-12pt}
	\label{fig:tpchresults}	
\end{figure} 
}

\noindent
\emph{4. Regression due to partial aggregation.}  For Q17, the learned cost models add local aggregation before the global one to reduce data transfer. However, this degrades the latency by $10\%$ and total processing time by $25\%$ as it does not help in reducing data. 
Currently, learned models do not learn from their own execution traces. We believe that doing so can potentially resolve some of the regressions.

\vspace{-0.20cm}
\subsubsection{Training and Runtime Overheads.}
\label{sec:overhead}
\rev{
We now describe the training and run-time overheads of \system. 
It takes less than 1 hour to analyze and learn models for a cluster running about $800$ jobs a day, and less than 4 hours for training over $50$K jobs instances at \microsoft{}. We use a parallel model trainer that leverages \scope{} to train and validate models independently and in parallel, which significantly speeds up the training process. For a single cluster of about $800$ jobs, \system learns about $23$K models which when simultaneously loaded takes about $600$ MB of memory. About $80$\% of the memory is taken by the individual models while the remaining is used by the combined model. The additional memory usage is not an issue for big data environments, where an optimizer can typically have $100$s of GBs of memory. Finally, we saw between $5$-$10$\% increase in the optimization time for most of the jobs when using learned cost models, which involves the overhead of signature computation of sub-graph models, invoking learned models, as well as any changes in plan exploration strategy due to learned models. Figure~\ref{fig:overheadtime} depicts the overhead in the optimization time for each of the $17$ production jobs we executed. Since the optimization time is often in orders of few hundreds of milliseconds, the overhead incurred here is negligible compared to the overall compilation time (orders of seconds) of jobs in big data systems as well as the potential gains we can achieve in the end-to-end latency (orders of 10s of minutes). 
}

\begin{figure}
	\papertext{\vspace{-0.15cm}}
	\centerline {
		\hbox{\resizebox{0.70\columnwidth}{!}{\includegraphics{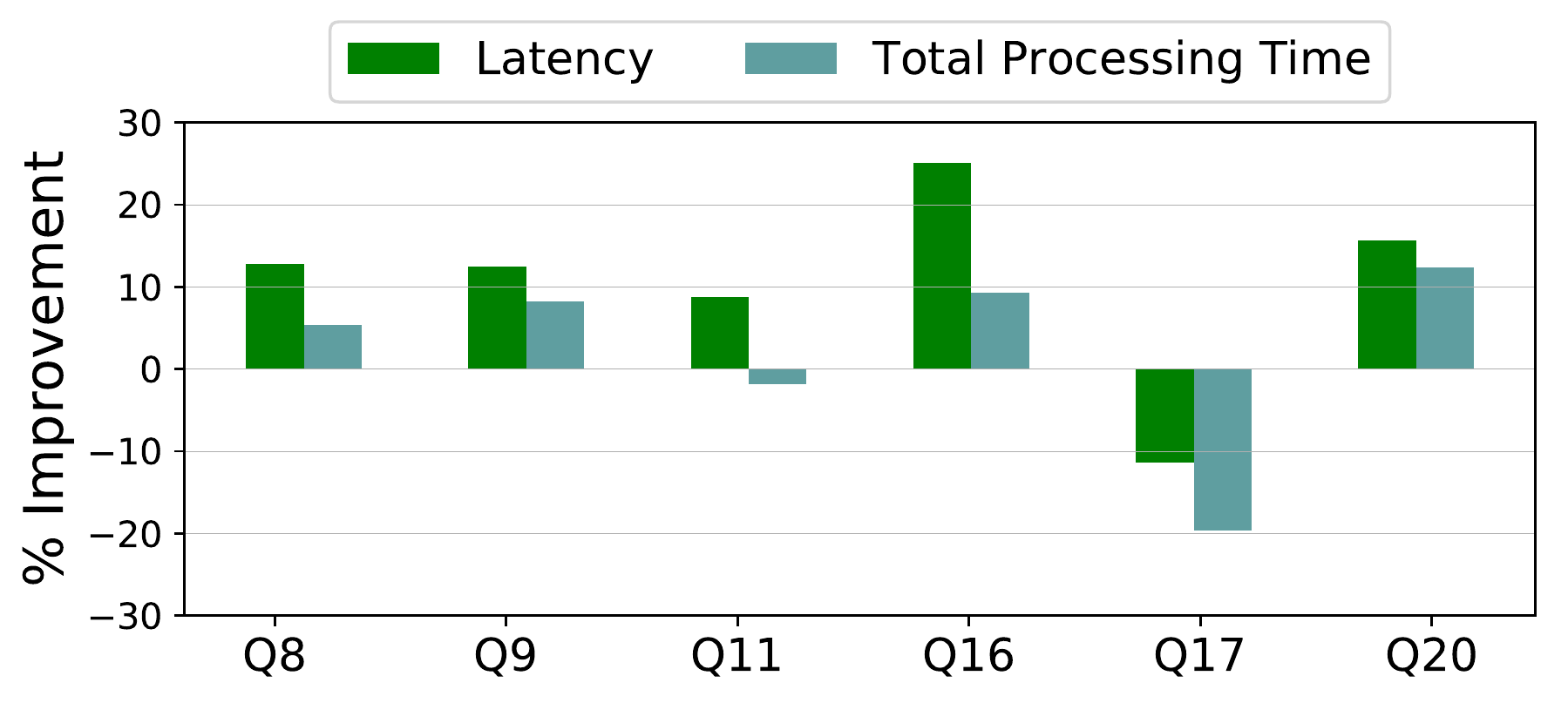}}}}
	\vspace{-0.4cm}
	\caption{Performance change with SCOPE + CLEO for TPC-H queries (higher is better)}
	\vspace{-0.55cm}
	\label{fig:tpchresults}
\end{figure}

\techreport{
\subsection{Discussion}
\label{sec:discussion}
We see that it is possible to learn accurate yet robust cost models from cloud workloads. 
Given the complexities and variance in the modern cloud environments, the model accuracies in \system are not perfect. Yet, they offer two to three orders of magnitude more accuracy and improvement in correlation from less than $0.1$ to greater than $0.7$ over the current state-of-the-art. 
The combined meta model further helps to achieve full workload coverage without sacrificing accuracy significantly. In fact, the combined model consistently retains the accuracy over a longer duration of time. 
While the accuracy and robustness gains from \system are obvious, the latency implications are more nuanced.
These are often due to various plan explorations made to work with the current cost models.
For instance, \scope{} jobs tend to over-partition at the leaf levels and leverage the massive scale-out possible for improving latency of jobs.

There are several ways to address performance regressions in production workloads. One option is to revisit the search and pruning strategies for plan exploration~\cite{cascades95} in the light of newer learned cost models. For instance, one problem we see is that current pruning strategies may sometimes skip operators without invoking their learned models. Additionally, we can also configure optimizer to not invoke learned models for specific operators or jobs.
Another improvement is to optimize a query twice (each taking only few orders of milliseconds), with and without \system, and select the plan with the better overall latency, as predicted using learned models since they are highly accurate and correlated to the runtime latency.
We can also monitor the performance of jobs in pre-production environment and isolate models that lead to performance regression (or poor latency predictions), and discard them from the feedback. This is possible since we do not learn a single global model in the first place. Furthermore, since learning cost models and feeding them back is a continuous process, regression causing models can self-correct by learning from future executions.
Finally, regressions for a few queries is not really a problem for ad-hoc workloads, since majority of the queries improve their latency anyways and reducing the overall processing time (and hence operational costs) is generally more important.


Finally, in this paper, we focused on the traditional use-case of a cost model for picking the physical query plan. However, several other cost model use-cases are relevant in cloud environments, where accuracy of predicted costs is crucial. Examples include 
performance prediction~\cite{perforator},
allocating resources to queries~\cite{jyothi2016morpheus},
estimating task runtimes for scheduling~\cite{boutin2014apollo},
estimating the progress of a query especially in server-less query processors~\cite{lee2016operator},
and running what-if analysis for physical design selection~\cite{chaudhuri2007self}.
Exploring these would be a part of future work.
Examples include 
performance prediction~\cite{perforator},
allocating resources to queries~\cite{jyothi2016morpheus},
estimating task runtimes for scheduling~\cite{boutin2014apollo},
estimating the progress of a query especially in server-less query processors~\cite{lee2016operator},
and running what-if analysis for physical design selection~\cite{chaudhuri2007self}.
Exploring these would be a part of future work.
}

\vspace{-0.2cm}
\section{Related Work}
\label{sec:related}
\vspace{-0.1cm}
Machine learning has been used for estimating the query execution time of a given physical plan in centralized databases~\cite{ganapathi2009predicting,akdere2012learning,li2012robust}.
In particular, the operator and sub-plan-level models in ~\cite{akdere2012learning} share similarities with our operator and operator-subgraph model. However, we discovered the coverage-accuracy gap between the two models to be substantially large. To bridge this gap, we proposed additional mutually enhancing models and then combined the predictions of these individual models to achieve the best of accuracy and coverage. 
There are other works on query progress indicators~\cite{chaudhuri2004estimating,luo2004toward} that use the run time statistics from the currently running query to tell how much percentage of the work has been completed for the query. Our approach, in contrast, uses compile time features to make the prediction before the execution starts.

Cardinality is a key input to cost estimation and several learning and feedback-driven approaches~\cite{cardLearner,rope,stillger2001leo}.
\rev{However, these works have either focused only on recurring or strict subgraphs~\cite{cardLearner, rope}, or learn only the ratio between the actual and predicted cardinality~\cite{stillger2001leo} that can go wrong in many situations, e.g., partial coverage results in erroneous estimates due to mixing of disparate models.
Most importantly, as we discuss in Section~\ref{sec:motivation}, fixing cardinalities alone do not always lead to accurate costs \rev{in big data systems}. There are other factors 
such as resources (e.g., partitions) consumed, operator implementations (e.g., custom operators), and hardware characteristics (e.g., parallel distributed environment) that could determine cost. In contrast to cardinality models, \system introduces novel learning techniques (e.g., multiple models, coupled with an ensemble) and extensions to optimizer to robustly model cost. That said, cardinality is still an important feature (see Figure~\ref{fig:peroperatorsubgraph}), and is also key to deciding partition counts, memory allocation at runtime, as well as for speculative execution in the job manager. A more detailed study on cardinality estimates in big data systems is an interesting avenue for future work.}
 
Several works find the optimal resources given a physical query execution plan~\cite{ernest,perforator,cherrypick}. They either train a performance model or apply non-parametric Bayesian optimization techniques with a few sample runs to find the optimal resource configuration.
However, the optimal execution plan may itself depend on the resources, and therefore in this work, we jointly find the optimal cost and resources. Nevertheless, the ideas from resource optimization work can be leveraged in our system to reduce the search space, especially if we consider multiple hardware types.

\techreport{Generating efficient combination of query plans and resources are also relevant to the new breed of serverless computing, where 
users are not required to provision resources explicitly and they are billed based on usage~\cite{serverless}.
For big data queries, this means that the optimizer needs to accurately estimate the cost of queries for given resources {\it and} explore different resource combinations so that users do not end up over-paying for their queries.}

Finally, several recent works apply machine learning techniques to improve different components of a data system~\cite{kraska2019sagedb,marcus2019neo}. The most prominent being learned indexes~\cite{kraska2018case}, which overfits a given stored data to a learned model that provides faster lookup as well as smaller storage footprint. 
~\cite{marcus2019neo} takes a more disruptive approach, where the vision is to replace the traditional query optimizers with one built using neural networks. 
In contrast, our focus in this paper is on improving the cost-estimation of operators in big data systems and our goal is to integrate learned models into existing query optimizers in a minimally invasive manner.

\vspace{-0.2cm}
\papertext{\vspace{-2pt}}
\section{Conclusion}
\label{sec:conclusion}
\vspace{-2pt}
Accurate cost prediction is critical for resource efficiency in \rev{big data} systems.
At the same time, modeling query costs is incredibly hard in \rev{these} systems.
In this paper, we present techniques to learn cost models from the massive cloud workloads.
We recognize that cloud workloads are highly heterogeneous in nature and {\it no one model fits all}. 
Instead, we leverage the common subexpression patterns in the workload and learn specialized models for each pattern. 
We further describe the accuracy and coverage trade-off of these specialized models and present additional mutual enhancing models to bridge the gap.
We combine the predictions from all of these individual models into a {\it robust model} that provides the best of both accuracy and coverage over a sufficiently long period of time.
A key part of our contribution is to integrate the learned cost models with existing query optimization frameworks. We present details on integration with \scope{}, a Cascade style query optimizer, and show how the learned models could be used to efficiently find both the query and resource optimal plans.
Overall, applying machine learning to systems is an active area of research, and this paper presents a practical approach for doing so deep within the core of a query processing system.


\balance


\small

\bibliographystyle{abbrv}
\bibliography{references}  

\end{document}